\documentclass[12pt,preprint]{aastex}

\usepackage{amsmath}
\usepackage{amssymb}
\usepackage{xspace}
\usepackage{color}
\usepackage{ifthen}
\usepackage{xspace}

\newcommand{\Msun}{{\ensuremath{\mathrm{M}_{\odot}}}}

\newcommand{\lSect}[1]{{\label{sec:#1}}}
\newcommand{\lFig}[1]{{\label{fig:#1}}}

\newcommand{\lTab}[1]{{\label{tab:#1}}}

\newcommand{\BE}{{\ensuremath{BE}}}
\newcommand{\Mrem}{{\ensuremath{M_{\mathrm{remnant}}}}}
\newcommand{\Mbar}{{\ensuremath{M_{\mathrm{baryon}}}}}
\newcommand{\Rrem}{{\ensuremath{R_{\mathrm{remnant}}}}}
\newcommand{\clight}{{\ensuremath{\mathrm{c}}}}
\newcommand{\G}{{\ensuremath{\mathrm{G}}}}

\newcommand{\BES}{{\ensuremath{BE_{\mathrm{S=4}}}}\xspace}
\newcommand{\MS}{{\ensuremath{M_{\mathrm{S=4}}}}\xspace}
\newcommand{\Reff}{{\ensuremath{R_{\mathrm{eff}}}}\xspace}
\newcommand{\BEYecore}{{\ensuremath{BE_{\mathrm{Ye\ core}}}}\xspace}
\newcommand{\BEFecore}{{\ensuremath{BE_{\mathrm{Fe\ core}}}}\xspace}
\newcommand{\Ye}{{\ensuremath{Y_e}}\xspace}
\newcommand{\Rsun}{{\ensuremath{\mathrm{R}_{\odot}}}\xspace}
\newcommand{\Mpresn}{{\ensuremath{M_{\mathrm{final}}}}\xspace}

\newcommand{\B}{{\ensuremath{\mathrm{B}}}\xspace}
\newcommand{\kB}{{\ensuremath{k_\mathrm{B}}}\xspace}

\def\gtaprx {\lower .1ex\hbox{\rlap{\raise .6ex\hbox{\hskip .3ex
	{\ifmmode{\scriptscriptstyle >}\else
		{$\scriptscriptstyle >$}\fi}}}
	\kern -.4ex{\ifmmode{\scriptscriptstyle \sim}\else
		{$\scriptscriptstyle\sim$}\fi}}}
\def\ltaprx {\lower .1ex\hbox{\rlap{\raise .6ex\hbox{\hskip .3ex
	{\ifmmode{\scriptscriptstyle <}\else
		{$\scriptscriptstyle <$}\fi}}}
	\kern -.4ex{\ifmmode{\scriptscriptstyle \sim}\else
		{$\scriptscriptstyle\sim$}\fi}}}
\newcommand{\FIGFF}[2]{{\ref{fig:#2}{#1}}}
\newcommand{\Figff}[1]{{\FIGFF{}{#1}}}
\newcommand{\FIG}[2]{{Fig.~\FIGFF{#1}{#2}}}
\newcommand{\Fig}[1]{{\FIG{}{#1}}}
\newcommand{\FIGS}[2]{{Figs.~\FIGFF{#1}{#2}}}
\newcommand{\Figs}[1]{{\FIGS{}{#1}}}

\newcommand{\TABFF}[1]{{\ref{tab:#1}}}
\newcommand{\Table}[1]{{Table~\TABFF{#1}}}
\newcommand{\Tables}[1]{{Tables~\TABFF{#1}}}
\newcommand{\Sectff}[1]{{\ref{sec:#1}}}
\newcommand{\Sect}[1]{{\S~\Sectff{#1}}}

\begin{document}

\shorttitle{Fallback in Supernovae}
\shortauthors{Zhang et al.}

\title{Fallback and Black Hole Production in Massive Stars} 

\author{Weiqun Zhang\altaffilmark{1}, S. E. Woosley\altaffilmark{2},
and A. Heger\altaffilmark{2,3}}

\altaffiltext{1}{Kavli Institute for Particle Astrophysics and Cosmology,
  Stanford University,
  P.O. Box 20450, MS 29, 
  Stanford, CA 94309; Chandra Fellow}
\altaffiltext{2}{Department of Astronomy and Astrophysics, University
of California, Santa Cruz, CA 95064}
\altaffiltext{3}{T6 Los Alamos National Laboratory, Los Alamos, New Mexico} 

\begin{abstract}
The compact remnants of core collapse supernovae - neutron stars and
black holes - have properties that reflect both the structure of their
stellar progenitors and the physics of the explosion. In particular,
the masses of these remnants are sensitive to the density structure of
the presupernova star and to the explosion energy.  To a considerable
extent, the final mass is determined by the ``fallback'', during the
explosion, of matter that initially moves outwards, yet ultimately
fails to escape. We consider here the simulated explosion of a large
number of massive stars (9 to 100 \Msun) of Population I (solar
metallicity) and III (zero metallicity), and find systematic
differences in the remnant mass distributions.  As pointed out by
\citet{Che89}, supernovae in more compact progenitor stars have
stronger reverse shocks and experience more fallback.  For Population
III stars above about 25 \Msun \ and explosion energies less than $1.5
\times 10^{51}$ erg, black holes are a common outcome, with masses
that increase with increasing main sequence mass up to a maximum hole
mass, for very low explosion energy, of about 40 \Msun.  If such stars
produce primary nitrogen, however, their black holes are
systematically smaller. For modern supernovae with nearly solar
metallicity, black hole production is much less frequent and the
typical masses, which depend sensitively on explosion energy, are
smaller. The maximum black hole mass is about 15 \Msun. We explore the
neutron star initial mass function for both populations and, for
reasonable assumptions about the initial mass cut of the explosion,
find good agreement with the average of observed masses of neutron
stars in binaries. We also find evidence for a bimodal distribution of
neutron star masses with a spike around 1.2 \Msun\ (gravitational
mass) and a broader distribution peaked around 1.4 \Msun.
\end{abstract}


\section{INTRODUCTION}
\lSect{intro}

\citet{Col71} first introduced the idea of fallback in supernovae,
attributing it to accretion in the rarefaction behind the outgoing
shock. \citet{Che89} discussed fallback in supernovae extensively and
emphasized that greater accretion would occur in compact progenitors.
For SN 1987A, a blue supergiant, Chevalier estimated a relatively large
fallback mass of $\sim$0.1 \Msun, and, for the more common Type II
supernovae from red supergiants, a value roughly 100 times smaller.
He also found, using self-similarity arguments, that the
accretion rate at late times when expansion dominated should scale as
$t^{-5/3}$, and emphasized the role of the reverse shock in
fallback \citep[see also][]{Col88}.  \citet{WW95} studied fallback
numerically in a variety of supernovae with different masses and
compositions, and emphasized black hole formation as an important
outcome for stars of higher mass and lower metallicity, with important
ramifications for their nucleosynthesis.  \citet{Mac01} studied
fallback numerically in a 25 \Msun \ supernova with varying explosion
energy and discussed the relevance of fallback for producing gamma-ray
bursts.

Thus far, however, there has been no systematic study of fallback in
stars with a very low metal content to determine the properties of
gravitational remnants that might have existed following a first
generation of stars. It has also been some time since the remnant
masses of solar metallicity stars were systematically explored
\citep{Tim96}, and no such studies have included the effects of mass
loss.  Calculations of fallback can be greatly influenced by the way
the inner boundary is handled \citep{Mac01}. This is particularly true
in cases where a piston or reflecting inner boundary has been used to
simulate the explosion and is still present in the calculation at late
times \citep[e.g.,][]{WW95}. As we shall see, for modern supernovae
that are red giants when they die, the error introduced by this
artificial inner boundary is small, but it can become appreciable for
zero metallicity stars with a much larger amount of fallback.
Since the material that falls back must be subtracted from the element
production for a given star, our results are also relevant for
calculations of nucleosynthesis and (radioactive-powered) light curves.

We do not study fallback in stars above 100 \Msun \ and leave out the
effects of rotation. Above 100 \Msun \ and below 260\,\Msun,
non-rotating stars encounter the pair instability and either lose
their outer layers before explosion (pulsational pair instability) or
explode completely without fallback. Above 260 \Msun, they collapse to
black holes \citep{HW02}. We also study here only single stars, not
binaries. The complications introduced by rotation and binary
membership could be included in future studies. A {\sl very}
approximate mapping between the results of binary and single star
evolution can be obtained by comparing two stars with the same final
helium core (or carbon-oxygen core) and explosion energy
\citep{Wel99,Fry02}. Core masses for zero metallicity and solar
metallicity stars are given here and in \citet{Woo02}. However, one of
our main results here is that the mass and radius of the hydrogen envelope
also greatly affect the fall back and therefore the remnant mass.  The
structure of the entire star must be considered, not just its core
mass. Similar caveats apply to the effects of rotation. Rotation tends
to increase the helium core mass and thus make larger black holes for
a given main sequence mass. However, a correct calculation of the
remnant mass involves not only the central engine (possibly affected by
rotation) and the core mass, but also the mass and radius of the hydrogen
envelope.

We also do not include in our study any asymptotic giant branch stars.
Stars less massive than about 9 \Msun\ develop cores of either carbon
and oxygen or neon, oxygen and magnesium that are increasingly
degenerate. The core mass is thus limited by the Chandrasekhar mass,
1.39 \Msun \ (for an electron mole number, $Y_e$ = 0.50). The fraction
of such stars that produce neutron stars is probably small for solar
metallicity \citep{pol07a}, but could be large at low metallicity
\citep{pol07b}.  It is uncertain what fraction, if any, of these stars
actually reach the Chandrasekhar mass without first losing their
envelope to winds and instabilities, and those with carbon-rich cores
will produce thermonuclear supernovae, not neutron stars. For any that
do make neutron stars, fall back is likely to be negligible and the
baryonic masses of remnants will be $\sim$1.39 \Msun. Correcting for
neutrino losses, the neutron star gravitational mass would be
$\sim$1.26 \Msun. Because of uncertain statistics, such neutron stars
are not included in our analysis, but could be by others.

\section{INITIAL MODELS}
\lSect{initmod}

The supernova models studied here are taken from two recent surveys by
\citet{HW07} and \citet{Woo07}. In each case, stars of various masses
and metallicities were evolved using the Kepler code
\citep{Wea78,Woo02} through all stable stages of nuclear burning until
their iron cores became unstable to collapse. The stars were then
exploded using pistons located at or near the edge of their iron
cores. For a discussion of how the piston was located and moved, and
for further details of these explosion models, see \citet{Woo02,Woo07}.

The first of these surveys examined the evolution and simulated
explosion of approximately 120 massive stars with masses in the range
10 - 100 \Msun \ and zero initial metallicity (hence Population III;
\Table{pre-SN-zero}). Heger and Woosley explored twelve different
choices of explosion energy and piston location for each mass.  While
results are given for all of them in the tables, the discussion here
focuses on just five.  The model names are given by a capital letter
``Z'', for ``zero'' metallicity, followed by a letter indicating the
piston location and explosion energy.  Four of these, Series ZB, ZD,
ZG, and ZJ, had the piston located at that point in the star where the
entropy equals 4.0\,\kB/baryon (typically this occurs at the base of
the oxygen burning shell) and kinetic energies of 0.6, 1.2, 2.4, and
10 B respectively (henceforth 1 B = 1 Bethe = 10$^{51}$\,erg). Series
P had the piston located deeper in, at the edge of the deleptonized
core (where $Y_e$, drops precipitously below 0.5 due to electron
capture) and had an explosion energy of 1.2 B.  Note that the
explosion energies quoted here are not the energy input by the piston,
but rather the kinetic energy of the ejecta at infinity.

The second survey treated a coarser grid of stellar masses (31 stars)
with solar metallicity and masses in the range 12 - 100\,\Msun\
(\Table{pre-SN-sollo03}). This survey is more appropriate to
supernovae today in the Milky Way Galaxy.  Greatest attention is paid
here to Series SA, which had the piston at the place in the star where
the entropy per baryon, $S/N_Ak$, equals 4.0 and an explosion energy
of 1.2 B. Except for metallicity effects then, series ZD and SA are
directly comparable.  Three other explosion models were also
considered: SB which had the piston at the entropy = 4.0 point but had
an explosion energy of 2.4 B; SC, with the piston at the edge of the
iron core (mass fraction of nuclei heavier than chromium greater than
50\%) and an explosion energy of 1.2 B; and SD, with the piston at the
edge of the iron core and an explosion energy of 2.4 B.

Models SC are thus the solar metallicity counterparts of Models ZP but
with a slight difference. The ZP models put the piston at the edge of
the deleptonized core; the SC models, at the edge of the iron core.
The difference between these two cores in a given model is usually
quite small and we do not think it has a major effect on the outcome.

\Tables{pre-SN-zero} and \TABFF{pre-SN-sollo03} give an overview of
these presupernova models. See also \citet{HW07} and \citet{Woo07}.
For the solar case, a few additional models were computed at low and
high mass using the same code and physics as the original surveys.
The tables give the initial mass of the star, its final mass (for the
Pop III stars, this is identical to the initial mass), the mass of the
location where an entropy of $S/N_Ak = 4.0$ is reached ($\MS$), the
size of the core where extensive electron capture has occurred ($\Ye$
core), the binding energies outside these two cores (binding energy
minus internal energy; \BEYecore, \BES), and the final radius of the
star (\Reff).  In the $\Reff$ column, ``WNL'' indicates a
hydrogen-rich Wolf-Rayet (WR) star and ``WC''/``WO'' indicate
early-type hydrogen-free WR stars.  Such WR stars have optically thick
winds with a photospheric ``effective radius'' located in this wind
regime.  Among the hydrogen-free WR stars, we found only the
carbon-rich and oxygen-rich subtypes (WC and WO) at presupernova, but
no early-type hydrogen-free WR stars that only display the pure
CNO-processed N-rich He layer (WNE stars).  There may be a very small
transition regime between 40 and 45\,\Msun where such WNE stars occur.

At 45\,\Msun, WO stars start to be produced as material from a late
helium burning stage in which oxygen dominates over carbon is exposed
to the surface.  At initial masses above $\sim60\,\Msun$ carbon
dominates over oxygen at the time the stars explodes. The final mass
of the star becomes smaller having lower WR mass loss rates at the end
and the stars lose mass from earlier phases of helium core
burning. Both effects increase the \emph{final} carbon-to-oxygen ratio
at the surface.

\section{CALCULATIONS}
\lSect{calculations}

Calculations were carried out using, Pangu, a one-dimensional
hydrodynamics code based on the second-order semi-discrete
finite-difference central scheme of \citet{KT00}.  Time evolution is
carried out by a third-order total variation diminishing Runge-Kutta
method \citep{SO89}.  We extended the scheme to spherical coordinates
based on the conservative form of hydrodynamics equations. The
treatment of spherical coordinates is the same as that in the RAM code
\citep{ram}.  In spherical coordinates, extra source terms are added
to the equations.  Geometric correction to the surface area and volume
of discretized numerical cells is applied when the numerical flux is
used to update conserved variables (density, momentum and total
energy) in the cells.

Gravity is implemented as source terms of the hydrodynamics equations.
A point mass is placed at the center of the grid.  The gravitational
force at a grid point is calculated from the enclosed mass, which
includes the central point mass and mass of the material on the
computational grid.  The central point mass is being updated by
keeping track of the mass flux across the inner boundary.

The supernova models were linked from the Kepler code, in which they
were initially calculated, to the Pangu code 100\,s after the shock
wave had been initiated. This typically corresponded to a time when
explosive nucleosynthesis had ended and the outgoing shock was just
exiting the core of helium and heavy elements, before it had
encountered any appreciable fraction of the hydrogen envelope. The
reverse shock had thus not yet developed and, for the explosion
energies considered, no fallback had yet occurred.

An outflow boundary condition was used at the inner boundary.  That is,
the ghost cells are simple duplicates of the first numerical cell on
the grid.  This type of boundary is very simple to implement.  A
potential problem of essentially any numerical boundary is small
errors at the boundary could accumulate and affect the calculation.
To avoid the problem, one should make sure that the flow across the
boundary is supersonic.  Thus the information at the boundary cannot
propagate outwards and affect the upstream fluid.  In our
calculations, the inner boundaries are chosen to be small enough to
ensure the supersonic condition.  However, it could be expensive to
use a very small radius for the inner boundary because of the
constrain of the Courant-Friedrichs-Lewy condition.  Fortunately, the
sound speed at the inner boundary is decreasing during fallback due to
the decrease of temperature, whereas the infall velocity is increasing
during fallback.  Therefore the sonic point is moving outwards over
the time.  

Calculations are performed in two steps to save computing time.  In
the first step, the numerical grid has an inner boundary at $r =
10^9\,\mathrm{cm}$, which is also the inner boundary of the initial
Kepler models, and an outer boundary at $r = 10^{14}\,\mathrm{cm}$.  A
logarithmic grid with 1000 zones is used for the $r$-direction.  The
region outside the star is filled with a low density medium with a
pressure of $p = 10\,\mathrm{dyn}\,\mathrm{cm}^{-2}$, a density of
$\rho = 10^{-12}\,\mathrm{g}\,\mathrm{cm}^{-3}$ and zero velocity.
The calculation is run to $t = 10^5\,\mathrm{s}$.  Then the model is
remapped to a new grid for the second step of calculations.  
For red giants in which the forward shock could have moved beyond the
outer boundary at $r = 10^{14}\,\mathrm{cm}$ already at $t =
10^5\,\mathrm{s}$, the link to the second step is at an earlier time
(e.g., $t = 5 \times 10^4\,\mathrm{s}$) so that the forward shock
still presents at the second step.  

The grid for the second phase of calculations also has 1000
logarithmic zones, but the boundaries are at $r =
10^{10}\,\mathrm{cm}$ and $r = 10^{16}\,\mathrm{cm}$.  Again the
outside medium is set to a constant state with a pressure of $p =
10\,\mathrm{dyn}\,\mathrm{cm}^{-2}$, a density of $\rho =
10^{-12}\,\mathrm{g}\,\mathrm{cm}^{-3}$ and zero velocity.  The second
step of the calculation is run to at least $t = 10^{6}\,\mathrm{s}$.
Then the simulation continues till the accretion rate is below
$10^{-8}\,\mathrm{M}_{\odot}\,\mathrm{s}^{-1}$ or it has reached $t =
2.0\times 10^{6}\,\mathrm{s}$.

\section{RESULTS}
\lSect{results}

\subsection{Fallback in Population III Supernovae}

Two distinguishing properties of evolved Pop III stars are that they
have lost little mass and also typically have more compact envelopes
than modern stars.  Most of them die as hot blue stars.  In some of
the more massive stars, however, penetration of the convective helium
burning core into the hydrogen envelope leads to the enrichment of the
latter with super-solar abundances of carbon and nitrogen.  Hydrogen
shell burning by the CNO cycle then expands the star to supergiant
proportions. Another special case is stars around $100\,\Msun$ which
begin to encounter the pulsational pair instability.  Strong pulses
lead to the ejection of the entire hydrogen envelope and even parts of
the helium core before the final core collapse (e.g.,
\citealt{HW02}). This weakens the reverse shock in such stars.

In the usual case, however, the explosion of Pop III stars is
accompanied by a stronger reverse shock and much more fallback than in
their solar counterparts.  Since mass loss is likely to be greatly
reduced in stars with no metals (Kudritzki 2002, Mokiem et al. 2007,
though see also Ekstr{\"o}m, Meynet, \& Maeder 2006), higher main
sequence mass implies a monotonically increasing helium core mass when
the star dies, and along with it the potential for making more massive
compact remnants, especially if the explosion energy is small.  This
is particularly interesting since several current simulations of
primordial star formation (e.g., \citealt{ON06}) predict rather high
initial masses for these first stars. While not studied here, it is
expected that still more massive stars (i.e., much above 100 \Msun),
will encounter an increasingly violent pair instability leading to the
complete disruption of the star and, eventually, above about 260
\Msun, the direct production of massive black holes without an initial
supernova explosion \citep{HW02}. These limiting masses would be
reduced by rotation.

\subsubsection{Hydrodynamics in a Representative Case}

\Fig{comphydro} shows the pressure, density, and velocity profiles at
100\,s, 200\,s, and 1000\,s as calculated in a typical Pop III model, Z25D,
using both Pangu and Kepler. Both the forward and reverse shocks are
clearly visible in the pressure and velocity plots. The reverse shock
forms as the expanding helium core runs into the star's hydrogen
envelope \citep[where the quantity $\rho r^3$ increases;][]{WW95} and
is decelerated. The hydrogen envelope in the presupernova star had its
base at $1.5 \times 10^{10}$ cm. With time the reverse shock moves
inwards in mass but outwards in radius. Starting at the edge of the
helium core at 7.6\,\Msun, by 1000\,s the reverse shock has moved into
3.3\,\Msun. The forward shock at this time is located at 19.19\,\Msun
\ and will shortly exit the star.

In the part of the star that is sonically disconnected from the
origin, the results of Kepler and Pangu are in very good agreement. As
time passes, however, there is an increasing discrepancy near the
origin where Pangu gives much higher collapse speeds than Kepler,
since the latter increasingly feels the effect of the reflecting inner
boundary held fixed at $1.0 \times 10^9\,\mathrm{cm}$. The inner
boundary in Pangu is also located at $1.0 \times 10^{9}\,\mathrm{cm}$,
but matter can flow through it without deceleration.  The sonic radius
at 1000 s is located at $3.27 \times 10^{10}\,\mathrm{cm}$ where the
sound speed is $488\,\mathrm{km}\,\mathrm{s}^{-1}$.

\Fig{mdot25} gives the accretion rate as a function of time calculated
by Pangu for this model. There are clearly four stages to the
accretion: 1) an early rapid accretion of material that failed to
achieve escape speed on the first try; 2) a decline in accretion rate
to an asymptotic dependence on t$^{-5/3}$ as appropriate for free
expansion \cite{Che89}; 3) a greatly enhanced fallback as the reverse
shock arrives at the core at $1.17\times 10^{4}\,\mathrm{s}$; and 4) a
final stage of free expansion.

The final value of the remnant masses from Pangu can be determined in
two ways.  After a sufficiently long time (i.e., a while after the
reverse shock has arrived at the center), the inner part of the
supernova will approach its asymptotic behavior.  Thus, the profiles
of pressure, density and velocity near the center are very simple for
the last dump of the simulation.  Both density and pressure have a
negative gradient.  The velocity is negative near the center and
increases monotonically outwards.  In the first method, a lower bound
and upper bound of the final remnant mass can be estimated from the
last dump.  All material with a negative velocity will fall into the
center.  This gives us a lower bound estimate of the final remnant
mass.  All material with a velocity larger than the escape velocity
will be able to escape.  This gives us an upper bound estimate of the
mass.  Our first estimate is the average of the two bounds.  

The second estimate is based on the asymptotic behavior of the
accretion rate, $\dot{M} \sim t^{-5/3}$.  Using the point mass and
accretion rate at the last dump of the simulation, we can get the
second estimate by a simple analytic integration.  For most models,
the two estimates are almost the same.  For example, the difference is
less than 0.01 \Msun\ in 958 out of 1440 Z-series and 123 out of 124
S-series models.  This gives us more confidence about our results.  In
principle, the two estimates should be identical provided that the
simulation is run long enough.  To determine which estimate is more
accurate, we did the two estimates using earlier dumps.  We found that
the second estimate was generally more accurate.  In this paper, we
will use the values of the second estimate. \Tables{remnant} and
\TABFF{remnant_soll03} show the results of the final remnant masses
calculated by Pangu.  In the end, Pangu gave a remnant mass of
$4.157\,\Msun$ for this star (Z25D) whereas the corresponding
calculation with Kepler gave $2.173\,\Msun$.

\subsubsection{Remnant Masses for the Pop III Survey}

\Fig{remnantz} and \Fig{remlow} give the remnant masses for the
Population III survey. Above about 35 \Msun\ the results are
influenced by the possibility of primary nitrogen production in the
star \citep{HW07}. For such massive stars, the entropy barrier
separating the outer extent of the convective core during helium
burning is not sufficient to prohibit mixing with the hydrogen
envelope with its very weak burning shell (this phenomenon does not
occur in non-rotating stars of solar metallicity). The mixing of
hydrogen and hot carbon leads to the production of nitrogen which is
convected throughout most of the envelope. With the new large CNO
abundance, nuclear energy generation is increased and the star
eventually expands to red supergiant proportions. Stars that do not
make nitrogen in this way stay compact. As \Fig{remnantz} shows, the
result is two branches of remnant masses.

\Figs{nstar04-0.6B}, \Figff{nstar04-1.2B}, and \Figff{nstar04-2.4B}
show the distribution of remnant masses for primordial supernovae with
explosion energies $0.6\,$B, $1.2\,$B, and $2.4\,$B for a piston
located at the $S/N_Ak = 4$ point.  \Fig{nstar0Fe} shows a similar
remnant mass distribution for a piston located at the edge of the
deleptonized core, for an explosion energy of $1.2\,$B.  The
systematics of these results are discussed in \Sect{rem}.

\subsection{Fallback in Population I Supernovae} 

Massive Pop I stars differ from Pop III stars in that they always develop
strong hydrogen burning shells and become red supergiants. Their
envelopes are thus, globally speaking, less tightly bound than in Pop
III stars, and also have different profiles of $\rho r^3$ as a
function of radius. Consequently, reverse shocks are weaker in red
supergiants, as noted by \citet{Che89} and \citet{WW95}, and their
remnant masses are smaller.  Above about 35 \Msun \ solar metallicity
stars lose their envelopes to winds during the red giant stage and
become Wolf-Rayet stars. The Wolf-Rayet stars lose further mass so
that, for example, a star with an initial mass of 100\,\Msun \ dies
with a mass of only 6\,\Msun.  Such light stars obviously cannot leave
behind very massive black holes and, in fact, tend to leave neutron
stars.

\Fig{remnantsol} shows the remnant masses expected for solar
metallicity.  These masses are influenced both by the decreased amount
of fallback that happens in the reverse shock in red supergiants and
by the mass loss before the explosion, especially above 40\,\Msun.

\Fig{nstarsol4} shows the distribution of neutron star
\emph{gravitational} masses for the solar metallicity survey. The
properties of these are sensitive to the placement of the piston as
well as its energy and the figure is for a piston location at the
$S/N_Ak = 4$ point near the base of the oxygen shell and an explosion
energy of $1.2\,$B.  The insert shows the distribution of
\emph{baryonic} masses of black holes, on a logarithmic scale on the
\textsl{x}-axis.  The main figure and the insert are normalized to add
up to 100\,\% together (all remnants; see caption of \Fig{nstarsol4}
for details).  \Fig{nstarsolFe} shows the same diagram of remnant mass
distribution for a piston located deeper in the star, at the edge of
the iron core. Lower explosion energies than 1.2 B are not considered
here, but, like the Pop III explosions, would give larger black hole
masses, up to approximately the mass of the helium core in the
presupernova star \citep{Woo02}.

\section{REMNANTS}
\lSect{rem}

\subsection{Gravitational and Baryonic Masses}

The fallback calculations described above and as summarized in Tables
\ref{tab:remnant} and \ref{tab:remnant_soll03} give the {\sl baryonic}
remnant masses.  For neutron stars especially, a significant fraction
of this mass becomes binding energy and is radiated away in the form
of neutrinos.  This fraction can be estimated if the binding energy of
the neutron star is known, but is dependent upon the nuclear equation
of state employed.  Here the estimate of \cite{LP02} is adopted:
\begin{equation}
\BE=\frac35\beta\left(1-\frac12\beta\right)^{\!-1},\qquad 
\beta=\frac{\G \Mrem}{\Rrem \clight^2}
\end{equation}
where $\G$ is the gravitational constant, $\Mrem$ the gravitational
mass of the remnant, $\Rrem$ the radius of the remnant, and $\clight$
the speed of light.  \citet{LP02} recommend a radius of $\sim 12\,$km.
This equation can then be solved to give a remnant mass as a function
of baryonic mass, $\Mbar$:
\begin{equation}
\Mrem=\Mbar \left(1+\frac35 \frac{\G \Mbar}{\Rrem \clight^2}\right)^{\!-1}.
\end{equation}
Here, two choices of maximum neutron star mass are employed, 1.7 \Msun
\ and 2.0 \Msun.  The limiting baryonic mass for which such heavy
neutron stars are made is then computed from
\begin{equation}
\Mbar=\Mrem \left(1-\frac35 \frac{\G \Mrem}{\Rrem \clight^2}\right)^{\!-1}.
\end{equation}
For example, a maximum gravitational mass of 2.0 \Msun \ implies a
maximum baryonic mass of $2.35\,\Msun$.  For baryonic masses above
that limit, a black hole forms. Here any effects due to rotation are
neglected.

Remnants that collapse to black holes may also lose an appreciable
fraction of their baryonic mass in the formation process, but unlike
neutron stars, that fraction depends not just on the final state but
on the formation process. If the black hole forms promptly from a big
collapsing core, bypassing any neutron star stage, and if the fallback
of matter contributing to its mass is small or essentially spherically
symmetric, very little rest mass is radiated away in form of
neutrinos.  The gravitational mass approximately equals the baryonic
mass.  On the other hand, one could first form a massive neutron star
that cools, radiating away approximately $20\,\%$ of its rest mass
before it collapses.  If the black hole is a rapidly rotating Kerr
black hole, the binding energy of the last stable orbit is $42.3\,\%$
of the rest mass.  If the disk is hot enough and not advection
dominated, this energy is radiated away.  Depending on the size of the
black hole, its rotation, and how much mass it accreted through a
cooling disk and at what rotation rate of the black hole that
occurred, the gravitational mass could be some $20\,\%$ to $40\,\%$
smaller than the baryonic mass. 

For simplicity here, we assume that the gravitational mass of any
black hole remnant equals the mass of the baryons that made it with no
correction for neutrino losses.  It should be kept in mind, however,
that this is actually an upper limit to the mass of the black hole.
Perhaps more realistically, the binding energy of the heaviest stable
neutron star, about 0.25 \Msun, should be subtracted from all our
black hole remnant masses, assuming that, along the way, each black
hole was formed from a protoneutron star that reached its maximum mass,
radiated its binding energy, and then collapsed. In the spirit of the
rest of the paper, all effects due to rotation are neglected.

\subsection{The Corrected Remnant Mass Distribution}

The distribution of remnant masses is obtained by linear interpolation
of the remnant masses among the different initial masses.  The result
is then integrated over a Salpeter initial mass function (IMF) with
exponent $-1.35$.  The resulting mapping into bins is exact.  A bin
width of $0.025\,\Msun$ is used.  The averages and standard deviations
(\Table{rem}) are computed from this distribution.  For the black
holes, the average logarithmic mass (geometric mean) is also given.
The column ``BH (\%)'' gives the fraction of remnants, from the mass
range considered, that are black holes. The fraction of neutron stars
is one minus that number.

To round out the table, remnant masses for main sequence stars lighter
than the 12\,\Msun \ considered by \cite{Woo07} and the 10 \Msun
\ considered by \cite{HW07} were estimated. Presupernova models of
resulting from 10 and 11 solar mass solar metallicity stars were
computed using the same physics and codes as described in the review.
Because such stars result in a degenerate core surrounded by thin
layers of heavy elements, it is reasonable to expect fallback to be
negligible in the explosion. The (baryonic) remnant masses were just
taken to be $S/N_A k$=4.0 masses of the presupernova stars,
$1.37\,\Msun$ for the $11\,\Msun$ star and $1.35\,\Msun$ for the
$10\,\Msun$.  The same 1.35 \Msun \ value was taken to characterize
all stars down to $9.1\,\Msun$, the assumed transition to
super-asymptotic giant branch (SAGB) stars \citep{pol07a}.  For the
piston located at the Fe core, a baryonic remnant mass of $1.32\,\Msun$ was
assumed for the $11\,\Msun$ star and lighter stars.

For the zero metallicity stars, the remnant characteristics of the
10\,\Msun\ were assumed to hold down to the SAGB limit for Z = 0
stars, taken here to be 9.5\,\Msun.

\section{DISCUSSION AND CONCLUSIONS}
\lSect{dis}

\Table{rem_org} gives the statistical characteristics of sets of
compact remnants extracted from an IMF-averaged distribution of
supernovae of the two populations. Here a Salpeter IMF is assumed over
the entire mass range examined, 9 $\ltaprx$ M/\Msun \ $\ltaprx$
100. The error bars represent a one-sigma deviation in the
distribution. Different choices for the IMF could be explored by
others using the values in Tables \ref{tab:remnant} and
\ref{tab:remnant_soll03}. For the black hole masses, the logarithmic
average as well as the arithmetic average might be of interest and
both are given. The statistical results depend not only on the physics
of the explosion (piston mass and energy), but also on the assumed
maximum mass of the neutron star. Obviously, the heavier that maximum
mass, the fewer the number of black holes.

In general, the observed trends follow expectation. More energetic
explosions eject more matter, experience less fallback, and make
lighter compact remnants. Even the lowest energy explosions
considered, 0.3 B, eject most of the hydrogen envelope of all Pop III
stars. Thus a supernova-like display can be expected in all cases -
though the event may be very faint if the radius is small and no
$^{56}$Ni is ejected \citep{HW07,Sca05}. The mass of the black hole in
these low energy explosions approaches that of the helium core of the
presupernova star (\Fig{remnantz}), e.g., $\sim10$ \Msun \ in a 25
\Msun \ supernova and $\sim$40 \Msun \ in a 100 \Msun \ star. The
average black hole mass from a generation of such zero-metal stars
ranges from about 6 to 10 \Msun \ if one excludes hyper-energetic
explosions (5 B and more) and very low energy ones. There is great
variation about this mean though, and hole masses up to 40\,\Msun
\ are possible. The fraction of black hole remnants is also
high, typically 20 - 50\% and possibly as great as 90\%. If modern
supernovae can be taken as a guide, the results for $S/N_Ak$ = 4, 1.2
B case (Model SA) may be most realistic \citep{Woo07}.

The fraction of remnants that are black holes is clearly smaller for
modern (i.e., solar metallicity) stars, and the average mass of those
holes is smaller. The actual value is sensitive to the values adopted
for the maximum neutron star mass and explosion energies, but
percentages in the range 10 - 25\% are reasonable. Explosion energies
as great as 2.4 B would probably give Type II supernova light curves
that are too bright \citep{Woo07}. Typical black hole masses are around
3 \Msun \ unless the explosion energy is very low.

Experimental estimates for the average black hole mass are hard to
find, and it must be kept in mind that accurate values for the black
hole mass can only come from binaries where the evolution might have
been influenced by mass exchange.  Rotation can also affect the
relation between helium core mass and main sequence mass and possibly
lead to larger black holes. There is also a predisposition to find
massive black holes since it is the mass that is taken as an indicator
that the object is not a neutron star.  Still it is interesting that
rather large values for black hole masses have been reported in
systems that presumably were not particularly metal poor
\citep{Rem06,Har07}. Either such systems have experienced an atypical
evolution (either of the black hole progenitor star or the black hole
itself after it was born) or the explosion energies are substantially
less than what one commonly takes for Type II supernovae.

Much better experimental calibrations are available for neutron star
masses, though one still must be concerned about the favored selection
of objects in close binary systems. The average neutron star masses
for solar metallicity stars in \Table{rem_org} range from 1.33 to 1.47
\Msun.  This is to be compared with, e.g., estimates by \citet{Tho99}
of $1.35 \pm 0.04$ \Msun \ for 21 radio pulsars. While the agreement
of the averages is impressive, it is also noteworthy that many neutron
stars in our calculated data set have masses outside this range. In
fact the lightest neutron star in our theoretical sample has a
gravitational mass of 1.16 \Msun \ for the $S=4$ set and 1.08\,\Msun \
for the iron core set. There are also numerous cases of neutron stars
with gravitational masses around the maximum mass limit.

Two major deficiencies of the current study is that it does not
include the effects of rotation or of binary interaction. The former
will tend to increase the mass of the remnants for a given main
sequence star since it leads to a larger helium core mass. The latter
may lead to reduced masses for remnants, especially if the parent star
loses its envelope early on to a companion and loses a lot more mass
as a Wolf-Rayet star. Both effects could be included in future
studies. It would also be useful to explore a wider range of explosion
energies for the solar metallicity stars. We plan such a survey, with
mass and energy resolution more like the Pop III survey presented
here, in the very near future. For now we note that the maximum mass
black hole expected, even for low energy explosions, is approximately
the mass of the heaviest helium core in a presupernova star, i.e., 15
\Msun \ for solar metallicity stars and 40 \Msun \ for zero metallicity
stars \citep{Woo02}. Low metallicity stars above 260 \Msun \ can make
heavier black holes \citep{HW02} and that threshold could be reduced
by rotation.

\acknowledgments 

WZ has been supported by NASA through Chandra Postdoctoral Fellowship
PF4-50036 awarded by the Chandra X-Ray Observatory Center.  
SW has been supported by the NSF (AST 0206111) and the DOE SciDAC
Program (DOE DE-FC-02-01ER41176 and DOE DE-FC-02-06ER41438).  
AH carried out this work under the auspices of the National Nuclear
Security Administration of the U.S.\ Department of Energy at Los
Alamos National Laboratory under Contract No.\ DE-AC52-06NA25396, and
was supported by the DOE Program for Scientific Discovery through
Advanced Computing (SciDAC; DE-FC02-01ER41176).

{}

\clearpage

%%%% tables

% ============
%   pre-SN data for znuc fallback calculations
% ============
\begin{table}
\caption{Summary of $Z=0$ Presupernova Model Data\lTab{pre-SN-zero}}
\scalebox{0.6}{
\begin{tabular}{rrrrrr|rrrrrr}
\hline
\hline
\noalign{\smallskip}
mass & \MS & \Ye core & \BEYecore & \BES & \Reff & mass & \MS & \Ye core & \BEYecore & \BES & \Reff \\
(\Msun) & (\Msun) & (\Msun) & (\B) & (\B) & (\Rsun) & (\Msun) & (\Msun) & (\Msun) & (\B) & (\B) & (\Rsun) \\
\noalign{\smallskip}
\hline
\noalign{\smallskip}
 10.0 &  1.28 &  1.27 &  0.09 &  0.09 &      62 &  18.7 &  1.55 &  1.41 &  0.66 &  0.48 &      10 \\
 10.2 &  1.38 &  1.18 &  0.27 &  0.04 &      38 &  18.8 &  1.57 &  1.42 &  0.69 &  0.50 &      11 \\
 10.4 &  1.32 &  1.18 &  0.34 &  0.11 &      34 &  18.9 &  1.63 &  1.47 &  0.76 &  0.56 &      11 \\
 10.5 &  1.41 &  1.20 &  0.34 &  0.07 &      27 &  19.0 &  1.63 &  1.44 &  0.80 &  0.57 &      11 \\
 10.6 &  1.40 &  1.20 &  0.30 &  0.06 &      21 &  19.2 &  1.59 &  1.44 &  0.72 &  0.53 &      10 \\
 10.7 &  1.41 &  1.19 &  0.35 &  0.08 &      20 &  19.4 &  1.56 &  1.44 &  0.70 &  0.54 &      10 \\
 10.8 &  1.34 &  1.17 &  0.39 &  0.13 &      19 &  19.6 &  1.63 &  1.45 &  0.79 &  0.57 &      11 \\
 10.9 &  1.43 &  1.25 &  0.27 &  0.08 &      17 &  19.8 &  1.61 &  1.43 &  0.79 &  0.58 &      10 \\
 11.0 &  1.42 &  1.33 &  0.23 &  0.12 &      15 &  20.0 &  1.46 &  1.46 &  0.62 &  0.61 &      13 \\
 11.1 &  1.31 &  1.27 &  0.22 &  0.14 &      18 &  20.5 &  1.64 &  1.46 &  0.79 &  0.56 &      13 \\
 11.2 &  1.35 &  1.19 &  0.37 &  0.14 &      14 &  21.0 &  1.50 &  1.49 &  0.71 &  0.70 &      10 \\
 11.3 &  1.47 &  1.18 &  0.43 &  0.11 &      14 &  21.5 &  1.61 &  1.45 &  0.80 &  0.59 &      14 \\
 11.4 &  1.48 &  1.22 &  0.43 &  0.16 &      18 &  22.0 &  1.52 &  1.36 &  0.92 &  0.72 &      11 \\
 11.5 &  1.35 &  1.35 &  0.15 &  0.15 &      13 &  22.5 &  1.49 &  1.43 &  0.68 &  0.58 &      11 \\
 11.6 &  1.34 &  1.34 &  0.16 &  0.16 &      12 &  23.0 &  1.63 &  1.46 &  0.90 &  0.68 &      11 \\
 11.7 &  1.38 &  1.23 &  0.41 &  0.17 &      13 &  23.5 &  1.92 &  1.58 &  1.19 &  0.87 &      12 \\
 11.8 &  1.49 &  1.24 &  0.40 &  0.16 &      12 &  24.0 &  2.07 &  1.64 &  1.34 &  0.98 &      12 \\
 11.9 &  1.54 &  1.26 &  0.34 &  0.13 &      11 &  24.5 &  2.20 &  1.67 &  1.47 &  1.07 &      13 \\
 12.0 &  1.30 &  1.26 &  0.23 &  0.15 &      12 &  25.0 &  2.17 &  1.59 &  1.43 &  1.02 &      19 \\
 12.2 &  1.51 &  1.26 &  0.44 &  0.19 &      14 &  25.5 &  1.87 &  1.62 &  1.08 &  0.82 &      14 \\
 12.4 &  1.46 &  1.31 &  0.44 &  0.24 &      10 &  26.0 &  1.74 &  1.53 &  1.15 &  0.90 &      15 \\
 12.6 &  1.50 &  1.23 &  0.49 &  0.20 &      10 &  26.5 &  1.80 &  1.54 &  1.19 &  0.90 &      16 \\
 12.8 &  1.41 &  1.31 &  0.38 &  0.21 &      10 &  27.0 &  1.73 &  1.52 &  1.12 &  0.89 &      18 \\
 13.0 &  1.40 &  1.37 &  0.25 &  0.21 &      19 &  27.5 &  1.59 &  1.46 &  1.14 &  0.96 &      16 \\
 13.2 &  1.54 &  1.31 &  0.43 &  0.23 &      10 &  28.0 &  1.60 &  1.46 &  1.06 &  0.88 &      21 \\
 13.4 &  1.57 &  1.35 &  0.43 &  0.21 &     9.1 &  28.5 &  1.62 &  1.43 &  1.22 &  0.98 &      19 \\
 13.6 &  1.42 &  1.41 &  0.27 &  0.27 &      10 &  29.0 &  1.72 &  1.49 &  1.26 &  1.01 &      15 \\
 13.8 &  1.45 &  1.37 &  0.44 &  0.32 &     9.0 &  29.5 &  1.70 &  1.45 &  1.29 &  1.00 &      15 \\
 14.0 &  1.57 &  1.37 &  0.52 &  0.29 &     9.0 &  30.0 &  1.75 &  1.50 &  1.24 &  0.97 &      20 \\
 14.2 &  1.58 &  1.38 &  0.53 &  0.30 &     9.0 &  30.5 &  1.77 &  1.51 &  1.36 &  1.09 &      14 \\
 14.4 &  1.62 &  1.39 &  0.56 &  0.31 &     9.2 &  31.0 &  1.84 &  1.54 &  1.46 &  1.15 &      15 \\
 14.6 &  1.56 &  1.40 &  0.55 &  0.36 &     9.0 &  31.5 &  1.93 &  1.58 &  1.56 &  1.24 &      18 \\
 14.8 &  1.56 &  1.41 &  0.55 &  0.37 &     9.0 &  32.0 &  1.94 &  1.57 &  1.60 &  1.27 &      15 \\
 15.0 &  1.43 &  1.28 &  0.53 &  0.30 &      10 &  32.5 &  1.98 &  1.59 &  1.65 &  1.30 &      18 \\
 15.2 &  1.45 &  1.33 &  0.49 &  0.32 &      10 &  33.0 &  2.08 &  1.63 &  1.77 &  1.41 &      16 \\
 15.4 &  1.43 &  1.31 &  0.50 &  0.31 &      10 &  33.5 &  2.12 &  1.64 &  1.79 &  1.42 &      24 \\
 15.6 &  1.46 &  1.36 &  0.52 &  0.38 &     8.9 &  34.0 &  2.12 &  1.64 &  1.85 &  1.48 &      16 \\
 15.8 &  1.55 &  1.36 &  0.62 &  0.40 &     8.7 &  34.5 &  2.19 &  1.65 &  1.91 &  1.52 &      19 \\
 16.0 &  1.58 &  1.41 &  0.62 &  0.42 &     8.9 &  35.0 &  2.24 &  1.66 &  1.95 &  1.56 &      21 \\
 16.2 &  1.61 &  1.42 &  0.66 &  0.44 &     8.9 &  36.0 &  2.33 &  1.80 &  2.06 &  1.73 &      17 \\
 16.4 &  1.63 &  1.45 &  0.67 &  0.46 &     8.8 &  37.0 &  2.25 &  1.82 &  2.20 &  1.96 &      18 \\
 16.6 &  1.63 &  1.44 &  0.68 &  0.47 &     9.1 &  38.0 &  2.23 &  1.66 &  1.87 &  1.49 &      59 \\
 16.8 &  1.74 &  1.32 &  0.84 &  0.48 &      10 &  39.0 &  2.24 &  1.78 &  1.76 &  1.44 &     739 \\
 17.0 &  1.76 &  1.35 &  0.87 &  0.52 &     9.0 &  40.0 &  2.16 &  1.88 &  2.60 &  2.50 &      23 \\
 17.1 &  1.77 &  1.37 &  0.86 &  0.53 &     8.9 &  41.0 &  2.27 &  1.85 &  2.31 &  2.10 &      50 \\
 17.2 &  1.74 &  1.34 &  0.87 &  0.52 &     9.1 &  42.0 &  2.24 &  1.93 &  2.49 &  2.35 &      30 \\
 17.3 &  1.82 &  1.39 &  0.84 &  0.53 &     9.0 &  43.0 &  1.97 &  1.75 &  2.82 &  2.79 &      26 \\
 17.4 &  1.50 &  1.37 &  0.63 &  0.45 &      10 &  44.0 &  1.64 &  1.64 &  2.86 &  2.86 &      23 \\
 17.5 &  1.82 &  1.40 &  0.85 &  0.54 &     9.1 &  45.0 &  2.20 &  1.91 &  2.30 &  2.18 &     896 \\
 17.6 &  1.87 &  1.46 &  0.85 &  0.59 &      10 &  50.0 &  2.34 &  1.82 &  2.08 &  1.80 &   2,020 \\
 17.7 &  1.73 &  1.58 &  0.64 &  0.52 &     9.1 &  55.0 &  1.91 &  1.91 &  2.82 &  2.82 &   2,048 \\
 17.8 &  1.83 &  1.40 &  0.91 &  0.59 &     9.1 &  60.0 &  1.91 &  1.91 &  3.21 &  3.21 &     150 \\
 17.9 &  1.84 &  1.41 &  0.87 &  0.58 &     9.3 &  65.0 &  1.97 &  1.95 &  3.16 &  3.15 &   1,830 \\
 18.0 &  1.49 &  1.38 &  0.55 &  0.40 &      26 &  70.0 &  2.18 &  1.96 &  3.87 &  3.72 &     184 \\
 18.1 &  1.53 &  1.39 &  0.65 &  0.46 &      11 &  75.0 &  2.15 &  2.04 &  3.71 &  3.63 &   2,305 \\
 18.2 &  1.54 &  1.41 &  0.84 &  0.70 &     9.4 &  80.0 &  2.26 &  2.14 &  3.88 &  3.81 &   2,334 \\
 18.3 &  1.70 &  1.43 &  0.88 &  0.69 &     9.5 &  85.0 &  2.42 &  2.03 &  4.17 &  4.05 &   2,526 \\
 18.4 &  1.51 &  1.40 &  0.48 &  0.33 &      51 &  90.0 &  2.40 &  1.54 &  4.11 &  3.92 &   2,648 \\
 18.5 &  1.55 &  1.41 &  0.68 &  0.50 &      11 &  95.0 &  2.53 &  2.04 &  4.26 &  4.11 &   1,214 \\
 18.6 &  1.51 &  1.41 &  0.57 &  0.42 &      22 & 100.0 &  2.02 &  1.44 &  3.34 &  2.92 &     1.3 \\
\noalign{\smallskip}
\hline
\end{tabular}
}
\end{table}

% ============
%   pre-SN data for sollo03 fallback calculations
% ============
\begin{table}
\caption{Summary of Solar Metallicity Presupernova Model Data\lTab{pre-SN-sollo03}}
\scalebox{0.6}{
\begin{tabular}{rrrrrrr|rrrrrrr}
\hline
\hline
\noalign{\smallskip}
mass & \Mpresn & \MS & Fe core & \BEFecore & \BES & \Reff & mass & \Mpresn & \MS & Fe core & \BEFecore & \BES & \Reff \\
(\Msun) & (\Msun) & (\Msun) & (\Msun) & (\B) & (\B) & (\Rsun) & (\Msun) & (\Msun) & (\Msun) & (\Msun) & (\B) & (\B) & (\Rsun) \\
\noalign{\smallskip}
\hline
\noalign{\smallskip}
 10.0 &  9.70 &  1.35 &  1.30 &  0.19 &  0.11 &     458 &  27.0 & 15.21 &  1.74 &  1.52 &  1.08 &  0.83 &   1,449 \\
 11.0 & 10.67 &  1.37 &  1.31 &  0.23 &  0.19 &     558 &  28.0 & 15.17 &  1.54 &  1.48 &  1.09 &  1.03 &   1,466 \\
 12.0 & 10.91 &  1.53 &  1.36 &  0.30 &  0.17 &     618 &  29.0 & 14.17 &  1.64 &  1.47 &  1.05 &  0.85 &   1,477 \\
 13.0 & 11.40 &  1.55 &  1.40 &  0.46 &  0.28 &     709 &  30.0 & 13.88 &  1.73 &  1.50 &  1.08 &  0.84 &   1,489 \\
 14.0 & 12.01 &  1.70 &  1.51 &  0.44 &  0.28 &     759 &  31.0 & 13.63 &  1.70 &  1.48 &  1.12 &  0.86 &   1,446 \\
 15.0 & 12.79 &  1.81 &  1.48 &  0.53 &  0.32 &     803 &  32.0 & 13.41 &  1.78 &  1.52 &  1.22 &  0.94 &   1,362 \\
 16.0 & 13.59 &  1.50 &  1.37 &  0.51 &  0.34 &     839 &  33.0 & 13.24 &  1.84 &  1.55 &  1.30 &  1.01 &   1,296 \\
 17.0 & 14.12 &  1.54 &  1.40 &  0.57 &  0.39 &     883 &  35.0 & 13.66 &  1.97 &  1.63 &  1.47 &  1.16 &     WNL \\
 18.0 & 14.82 &  1.89 &  1.49 &  0.70 &  0.37 &     942 &  40.0 & 15.34 &  2.34 &  1.82 &  1.93 &  1.61 &     WNL \\
 19.0 & 15.48 &  1.64 &  1.45 &  0.68 &  0.45 &     990 &  45.0 & 13.02 &  2.27 &  1.79 &  1.76 &  1.44 &      WO \\
 20.0 & 15.93 &  1.82 &  1.54 &  0.89 &  0.60 &   1,032 &  50.0 &  9.82 &  1.70 &  1.49 &  1.05 &  0.81 &      WO \\
 21.0 & 16.16 &  1.46 &  1.46 &  0.48 &  0.47 &   1,085 &  55.0 &  9.38 &  1.65 &  1.47 &  1.03 &  0.82 &      WO \\
 22.0 & 16.16 &  1.84 &  1.54 &  0.95 &  0.65 &   1,139 &  60.0 &  7.29 &  1.60 &  1.45 &  0.71 &  0.53 &      WO \\
 23.0 & 16.37 &  2.12 &  1.73 &  1.18 &  0.86 &   1,207 &  70.0 &  6.41 &  1.72 &  1.50 &  0.82 &  0.56 &      WC \\
 24.0 & 16.22 &  2.05 &  1.70 &  1.17 &  0.87 &   1,270 &  80.0 &  6.37 &  1.66 &  1.48 &  0.76 &  0.54 &      WC \\
 25.0 & 15.84 &  1.90 &  1.59 &  1.16 &  0.86 &   1,329 & 100.0 &  6.04 &  1.81 &  1.54 &  0.81 &  0.58 &      WC \\
 26.0 & 15.41 &  1.73 &  1.54 &  0.97 &  0.74 &   1,386 & 120.0 &  6.00 &  1.60 &  1.43 &  0.68 &  0.48 &      WC \\
\noalign{\smallskip}
\hline
\end{tabular}
}
\end{table}

% ============
%   Z=0 Remnat Data
% ============
\begin{table}
\caption{$Z=0$ Baryonic Remnant Masses \lTab{remnant}}
\scalebox{0.6}{
\begin{tabular}{rrrrrrrrrrrrr}
\hline
\hline
\noalign{\smallskip}
Run & ZA & ZB & ZC & ZD & ZE & ZF & ZG & ZH & ZI & ZJ & ZP & ZV \\
Energy (\B) & 0.3 & 0.6 & 0.9 & 1.2 & 1.5 & 1.8 & 2.4 & 3.0 & 5.0 & 10.0 & 1.2 & 10.0 \\
Piston & $S=4$ & $S=4$ & $S=4$ & $S=4$ & $S=4$ & $S=4$ & $S=4$ & $S=4$ & $S=4$ & $S=4$ & Ye core & Ye core \\
\noalign{\smallskip}
\hline
\noalign{\smallskip}
Initial Mass & \multicolumn{12}{c}{\ \hrulefill\ Remnant Mass\ \hrulefill\ \ }\\
(\Msun) & (\Msun) & (\Msun) & (\Msun) & (\Msun) & (\Msun) & (\Msun) & (\Msun) & (\Msun) & (\Msun) & (\Msun) & (\Msun) & (\Msun) \\
\noalign{\smallskip}
\hline
\noalign{\smallskip}
 10.0 &  1.37 &  1.28 &  1.28 &  1.28 &  1.28 &  1.28 &  1.28 &  1.28 &  1.27 &  1.27 &  1.28 &  1.27 \\
 10.2 &  1.39 &  1.38 &  1.38 &  1.38 &  1.38 &  1.38 &  1.38 &  1.38 &  1.38 &  1.38 &  1.18 &  1.18 \\
 10.4 &  1.60 &  1.33 &  1.32 &  1.32 &  1.32 &  1.32 &  1.32 &  1.32 &  1.32 &  1.32 &  1.20 &  1.18 \\
 10.5 &  1.53 &  1.43 &  1.42 &  1.42 &  1.41 &  1.41 &  1.41 &  1.41 &  1.41 &  1.41 &  1.20 &  1.20 \\
 10.6 &  1.44 &  1.41 &  1.41 &  1.40 &  1.40 &  1.40 &  1.40 &  1.40 &  1.40 &  1.40 &  1.20 &  1.20 \\
 10.7 &  1.55 &  1.43 &  1.42 &  1.42 &  1.41 &  1.41 &  1.41 &  1.41 &  1.41 &  1.41 &  1.20 &  1.19 \\
 10.8 &  1.60 &  1.49 &  1.36 &  1.35 &  1.34 &  1.34 &  1.34 &  1.34 &  1.34 &  1.34 &  1.18 &  1.17 \\
 10.9 &  1.59 &  1.46 &  1.45 &  1.44 &  1.43 &  1.43 &  1.43 &  1.43 &  1.43 &  1.43 &  1.27 &  1.25 \\
 11.0 &  1.64 &  1.57 &  1.47 &  1.44 &  1.43 &  1.43 &  1.43 &  1.43 &  1.43 &  1.43 &  1.35 &  1.33 \\
 11.1 &  1.96 &  1.40 &  1.32 &  1.31 &  1.31 &  1.31 &  1.31 &  1.31 &  1.31 &  1.31 &  1.28 &  1.27 \\
 11.2 &  1.71 &  1.60 &  1.39 &  1.36 &  1.36 &  1.35 &  1.35 &  1.35 &  1.35 &  1.35 &  1.23 &  1.19 \\
 11.3 &  1.76 &  1.51 &  1.49 &  1.48 &  1.47 &  1.47 &  1.47 &  1.47 &  1.47 &  1.47 &  1.26 &  1.18 \\
 11.4 &  2.03 &  1.74 &  1.53 &  1.50 &  1.49 &  1.49 &  1.49 &  1.48 &  1.48 &  1.48 &  1.27 &  1.22 \\
 11.5 &  1.80 &  1.64 &  1.40 &  1.36 &  1.36 &  1.36 &  1.35 &  1.35 &  1.35 &  1.35 &  1.36 &  1.35 \\
 11.6 &  1.89 &  1.67 &  1.41 &  1.36 &  1.35 &  1.35 &  1.35 &  1.34 &  1.34 &  1.34 &  1.36 &  1.34 \\
 11.7 &  1.93 &  1.72 &  1.52 &  1.42 &  1.39 &  1.39 &  1.38 &  1.38 &  1.38 &  1.38 &  1.54 &  1.23 \\
 11.8 &  2.03 &  1.78 &  1.57 &  1.53 &  1.50 &  1.50 &  1.50 &  1.49 &  1.49 &  1.49 &  1.54 &  1.24 \\
 11.9 &  2.03 &  1.67 &  1.56 &  1.54 &  1.54 &  1.54 &  1.54 &  1.54 &  1.53 &  1.53 &  1.46 &  1.26 \\
 12.0 &  2.02 &  1.63 &  1.32 &  1.31 &  1.31 &  1.31 &  1.30 &  1.30 &  1.30 &  1.30 &  1.32 &  1.26 \\
 12.2 &  2.43 &  2.01 &  1.64 &  1.54 &  1.52 &  1.51 &  1.51 &  1.51 &  1.51 &  1.51 &  1.72 &  1.26 \\
 12.4 &  2.36 &  2.03 &  1.87 &  1.62 &  1.50 &  1.47 &  1.46 &  1.46 &  1.46 &  1.46 &  1.78 &  1.31 \\
 12.6 &  2.46 &  2.04 &  1.72 &  1.55 &  1.53 &  1.51 &  1.50 &  1.50 &  1.50 &  1.50 &  1.82 &  1.23 \\
 12.8 &  2.52 &  2.11 &  1.74 &  1.44 &  1.42 &  1.41 &  1.41 &  1.41 &  1.41 &  1.41 &  1.60 &  1.31 \\
 13.0 &  2.57 &  2.09 &  1.60 &  1.41 &  1.40 &  1.40 &  1.40 &  1.40 &  1.40 &  1.40 &  1.45 &  1.37 \\
 13.2 &  2.77 &  2.23 &  1.89 &  1.60 &  1.57 &  1.55 &  1.54 &  1.54 &  1.54 &  1.54 &  1.93 &  1.31 \\
 13.4 &  2.84 &  2.23 &  1.66 &  1.59 &  1.58 &  1.57 &  1.57 &  1.57 &  1.57 &  1.57 &  1.93 &  1.35 \\
 13.6 &  2.94 &  2.38 &  2.08 &  1.62 &  1.45 &  1.43 &  1.42 &  1.42 &  1.42 &  1.42 &  1.66 &  1.41 \\
 13.8 &  3.10 &  2.52 &  2.29 &  1.90 &  1.57 &  1.47 &  1.45 &  1.45 &  1.45 &  1.45 &  2.05 &  1.37 \\
 14.0 &  3.24 &  2.60 &  2.30 &  1.83 &  1.66 &  1.61 &  1.58 &  1.57 &  1.57 &  1.57 &  2.19 &  1.37 \\
 14.2 &  3.32 &  2.67 &  2.35 &  1.89 &  1.68 &  1.63 &  1.59 &  1.59 &  1.58 &  1.58 &  2.24 &  1.38 \\
 14.4 &  3.51 &  2.87 &  2.52 &  1.98 &  1.73 &  1.67 &  1.63 &  1.63 &  1.62 &  1.62 &  2.37 &  1.39 \\
 14.6 &  3.65 &  2.94 &  2.69 &  2.36 &  1.84 &  1.62 &  1.56 &  1.56 &  1.56 &  1.56 &  2.48 &  1.40 \\
 14.8 &  3.75 &  2.98 &  2.71 &  2.41 &  1.91 &  1.65 &  1.57 &  1.57 &  1.56 &  1.56 &  2.51 &  1.41 \\
 15.0 &  3.82 &  3.00 &  2.41 &  1.65 &  1.47 &  1.44 &  1.44 &  1.43 &  1.43 &  1.43 &  2.07 &  1.28 \\
 15.2 &  4.04 &  3.13 &  2.61 &  1.78 &  1.50 &  1.46 &  1.45 &  1.45 &  1.45 &  1.45 &  2.08 &  1.33 \\
 15.4 &  3.96 &  3.05 &  2.55 &  1.79 &  1.50 &  1.45 &  1.44 &  1.43 &  1.43 &  1.43 &  2.17 &  1.31 \\
 15.6 &  4.30 &  3.30 &  2.85 &  2.15 &  1.69 &  1.52 &  1.47 &  1.46 &  1.46 &  1.46 &  2.39 &  1.36 \\
 15.8 &  4.37 &  3.40 &  2.99 &  2.58 &  1.90 &  1.68 &  1.57 &  1.56 &  1.55 &  1.55 &  2.73 &  1.36 \\
 16.0 &  4.59 &  3.56 &  3.19 &  2.88 &  2.17 &  1.68 &  1.59 &  1.58 &  1.58 &  1.58 &  2.91 &  1.41 \\
 16.2 &  4.77 &  3.63 &  3.25 &  3.02 &  2.61 &  1.90 &  1.62 &  1.61 &  1.61 &  1.61 &  3.02 &  1.42 \\
 16.4 &  4.94 &  3.79 &  3.41 &  3.16 &  2.75 &  1.95 &  1.64 &  1.63 &  1.63 &  1.63 &  3.15 &  1.45 \\
 16.6 &  5.08 &  3.99 &  3.57 &  3.24 &  2.49 &  1.78 &  1.63 &  1.63 &  1.63 &  1.63 &  3.23 &  1.44 \\
 16.8 &  5.14 &  3.91 &  3.44 &  3.18 &  2.88 &  2.31 &  1.83 &  1.77 &  1.75 &  1.75 &  3.20 &  1.32 \\
 17.0 &  5.55 &  4.15 &  3.68 &  3.41 &  3.13 &  2.51 &  1.83 &  1.77 &  1.76 &  1.76 &  3.46 &  1.35 \\
 17.1 &  5.52 &  4.21 &  3.76 &  3.48 &  3.19 &  2.54 &  1.84 &  1.78 &  1.77 &  1.77 &  3.51 &  1.37 \\
 17.2 &  5.51 &  4.19 &  3.72 &  3.46 &  3.16 &  2.47 &  1.78 &  1.75 &  1.74 &  1.74 &  3.49 &  1.34 \\
 17.3 &  5.57 &  4.27 &  3.82 &  3.55 &  3.24 &  2.67 &  1.91 &  1.83 &  1.82 &  1.82 &  3.51 &  1.39 \\
 17.4 &  5.49 &  4.08 &  3.60 &  3.25 &  2.17 &  1.52 &  1.50 &  1.50 &  1.50 &  1.50 &  3.22 &  1.37 \\
 17.5 &  5.76 &  4.40 &  3.92 &  3.62 &  3.34 &  2.71 &  1.91 &  1.84 &  1.82 &  1.82 &  3.63 &  1.40 \\
 17.6 &  6.04 &  4.66 &  4.19 &  3.77 &  3.28 &  2.34 &  1.92 &  1.88 &  1.87 &  1.87 &  3.78 &  1.46 \\
 17.7 &  5.83 &  4.47 &  3.97 &  3.59 &  3.01 &  2.04 &  1.77 &  1.74 &  1.73 &  1.73 &  3.56 &  1.58 \\
 17.8 &  6.05 &  4.62 &  4.14 &  3.84 &  3.46 &  2.79 &  1.91 &  1.84 &  1.83 &  1.83 &  3.86 &  1.40 \\
 17.9 &  6.06 &  4.58 &  4.07 &  3.78 &  3.45 &  2.90 &  1.94 &  1.85 &  1.84 &  1.84 &  3.77 &  1.42 \\
 18.0 &  5.71 &  4.30 &  3.69 &  2.09 &  1.52 &  1.50 &  1.49 &  1.49 &  1.49 &  1.49 &  3.02 &  1.38 \\
 18.1 &  6.08 &  4.33 &  3.79 &  3.23 &  1.73 &  1.54 &  1.53 &  1.53 &  1.53 &  1.52 &  3.20 &  1.39 \\
 18.2 &  6.59 &  4.77 &  4.13 &  3.77 &  3.23 &  2.67 &  1.61 &  1.56 &  1.54 &  1.54 &  3.82 &  1.41 \\
 18.3 &  6.61 &  4.86 &  4.29 &  3.90 &  3.48 &  2.88 &  1.80 &  1.73 &  1.71 &  1.70 &  3.97 &  1.43 \\
 18.4 &  5.85 &  4.25 &  1.56 &  1.51 &  1.51 &  1.51 &  1.51 &  1.51 &  1.51 &  1.51 &  1.91 &  1.40 \\
 18.5 &  6.45 &  4.59 &  3.99 &  3.41 &  1.97 &  1.57 &  1.55 &  1.55 &  1.55 &  1.55 &  3.43 &  1.42 \\
\multicolumn{13}{r}{(continued on next page)}\\
\noalign{\smallskip}
\hline
\end{tabular}
}
\end{table}
\addtocounter{table}{-1}
\begin{table}
\caption{$Z=0$ Baryonic Remnant Masses (continued)}
\scalebox{0.6}{
\begin{tabular}{rrrrrrrrrrrrr}
\hline
\hline
\noalign{\smallskip}
Run & ZA & ZB & ZC & ZD & ZE & ZF & ZG & ZH & ZI & ZJ & ZP & ZV \\
Energy (\B) & 0.3 & 0.6 & 0.9 & 1.2 & 1.5 & 1.8 & 2.4 & 3.0 & 5.0 & 10.0 & 1.2 & 10.0 \\
Piston & $S=4$ & $S=4$ & $S=4$ & $S=4$ & $S=4$ & $S=4$ & $S=4$ & $S=4$ & $S=4$ & $S=4$ & Ye core & Ye core \\
\noalign{\smallskip}
\hline
\noalign{\smallskip}
Initial Mass & \multicolumn{12}{c}{\ \hrulefill\ Remnant Mass\ \hrulefill\ \ }\\
(\Msun) & (\Msun) & (\Msun) & (\Msun) & (\Msun) & (\Msun) & (\Msun) & (\Msun) & (\Msun) & (\Msun) & (\Msun) & (\Msun) & (\Msun) \\
\noalign{\smallskip}
\hline
\noalign{\smallskip}
 18.6 &  6.25 &  4.57 &  3.93 &  2.54 &  1.55 &  1.52 &  1.51 &  1.51 &  1.51 &  1.51 &  3.24 &  1.41 \\
 18.7 &  6.58 &  4.76 &  4.17 &  3.64 &  1.69 &  1.56 &  1.55 &  1.55 &  1.55 &  1.55 &  3.63 &  1.42 \\
 18.8 &  6.66 &  4.71 &  4.06 &  3.45 &  2.10 &  1.59 &  1.57 &  1.57 &  1.57 &  1.57 &  3.49 &  1.42 \\
 18.9 &  6.97 &  5.13 &  4.46 &  3.74 &  1.93 &  1.65 &  1.63 &  1.63 &  1.63 &  1.63 &  3.80 &  1.47 \\
 19.0 &  6.99 &  5.16 &  4.43 &  3.38 &  1.84 &  1.65 &  1.63 &  1.63 &  1.63 &  1.63 &  3.58 &  1.44 \\
 19.2 &  7.08 &  5.03 &  4.37 &  3.77 &  2.51 &  1.61 &  1.59 &  1.59 &  1.59 &  1.59 &  3.81 &  1.44 \\
 19.4 &  7.25 &  5.09 &  4.38 &  3.79 &  2.73 &  1.61 &  1.57 &  1.56 &  1.56 &  1.56 &  3.84 &  1.44 \\
 19.6 &  7.40 &  5.25 &  4.56 &  4.02 &  3.12 &  1.69 &  1.64 &  1.63 &  1.63 &  1.63 &  4.08 &  1.45 \\
 19.8 &  7.66 &  5.49 &  4.72 &  3.79 &  1.99 &  1.63 &  1.61 &  1.61 &  1.61 &  1.61 &  3.97 &  1.43 \\
 20.0 &  7.77 &  5.37 &  4.37 &  2.43 &  1.70 &  1.50 &  1.47 &  1.47 &  1.46 &  1.46 &  2.49 &  1.46 \\
 20.5 &  7.75 &  5.47 &  4.67 &  3.57 &  1.83 &  1.67 &  1.65 &  1.65 &  1.64 &  1.64 &  3.80 &  1.46 \\
 21.0 &  9.14 &  6.36 &  5.45 &  4.21 &  1.90 &  1.53 &  1.50 &  1.50 &  1.50 &  1.50 &  4.30 &  1.49 \\
 21.5 &  7.88 &  5.92 &  5.05 &  3.30 &  1.66 &  1.63 &  1.61 &  1.61 &  1.61 &  1.61 &  3.61 &  1.45 \\
 22.0 &  9.92 &  6.93 &  5.87 &  3.71 &  1.60 &  1.53 &  1.52 &  1.52 &  1.52 &  1.52 &  4.52 &  1.36 \\
 22.5 &  9.81 &  6.96 &  5.68 &  2.82 &  1.55 &  1.50 &  1.50 &  1.49 &  1.49 &  1.49 &  3.32 &  1.43 \\
 23.0 & 10.49 &  7.36 &  6.26 &  4.51 &  1.94 &  1.66 &  1.64 &  1.64 &  1.64 &  1.64 &  5.00 &  1.46 \\
 23.5 & 11.42 &  8.10 &  6.86 &  5.78 &  3.11 &  2.18 &  1.95 &  1.93 &  1.92 &  1.92 &  6.22 &  1.59 \\
 24.0 & 12.32 &  8.54 &  7.23 &  6.47 &  5.11 &  3.02 &  2.18 &  2.10 &  2.08 &  2.07 &  6.67 &  1.65 \\
 24.5 & 12.64 &  8.91 &  7.47 &  6.66 &  4.70 &  2.60 &  2.30 &  2.23 &  2.21 &  2.20 &  6.87 &  1.68 \\
 25.0 & 10.19 &  7.96 &  7.13 &  4.16 &  2.57 &  2.35 &  2.21 &  2.19 &  2.17 &  2.17 &  5.96 &  1.60 \\
 25.5 & 13.48 &  9.41 &  7.68 &  2.01 &  1.90 &  1.88 &  1.87 &  1.87 &  1.87 &  1.87 &  3.00 &  1.62 \\
 26.0 & 14.22 &  9.94 &  8.14 &  2.08 &  1.77 &  1.75 &  1.74 &  1.74 &  1.74 &  1.74 &  3.91 &  1.53 \\
 26.5 & 14.41 &  9.97 &  8.29 &  6.41 &  1.88 &  1.82 &  1.81 &  1.81 &  1.80 &  1.80 &  7.23 &  1.55 \\
 27.0 & 12.03 &  8.92 &  7.26 &  1.96 &  1.77 &  1.74 &  1.73 &  1.73 &  1.73 &  1.73 &  2.30 &  1.52 \\
 27.5 & 15.71 & 11.13 &  9.15 &  3.16 &  1.63 &  1.61 &  1.60 &  1.59 &  1.59 &  1.59 &  6.89 &  1.46 \\
 28.0 & 12.90 &  9.51 &  7.76 &  1.92 &  1.63 &  1.61 &  1.60 &  1.60 &  1.60 &  1.60 &  2.19 &  1.46 \\
 28.5 & 16.36 & 11.72 &  9.42 &  2.54 &  1.68 &  1.64 &  1.63 &  1.63 &  1.63 &  1.62 &  7.50 &  1.43 \\
 29.0 & 17.29 & 12.15 & 10.16 &  8.23 &  2.21 &  1.83 &  1.74 &  1.73 &  1.72 &  1.72 &  8.95 &  1.49 \\
 29.5 & 17.77 & 12.75 & 10.25 &  7.56 &  1.99 &  1.76 &  1.71 &  1.71 &  1.70 &  1.70 &  8.83 &  1.46 \\
 30.0 & 15.40 & 11.12 &  9.55 &  2.73 &  1.97 &  1.83 &  1.77 &  1.76 &  1.75 &  1.75 &  4.56 &  1.51 \\
 30.5 & 18.91 & 13.70 & 11.16 &  9.74 &  3.15 &  1.98 &  1.79 &  1.78 &  1.77 &  1.77 &  9.96 &  1.52 \\
 31.0 & 19.37 & 14.10 & 11.47 & 10.06 &  3.89 &  2.16 &  1.89 &  1.86 &  1.85 &  1.84 & 10.17 &  1.54 \\
 31.5 & 19.53 & 14.48 & 12.04 & 10.29 &  2.62 &  2.14 &  1.98 &  1.95 &  1.94 &  1.94 & 10.74 &  1.59 \\
 32.0 & 20.62 & 15.25 & 12.45 & 11.04 &  7.66 &  2.53 &  2.02 &  1.96 &  1.95 &  1.94 & 11.20 &  1.58 \\
 32.5 & 21.01 & 15.41 & 12.73 & 11.32 &  5.49 &  2.42 &  2.05 &  2.01 &  1.99 &  1.98 & 11.48 &  1.61 \\
 33.0 & 21.70 & 16.32 & 13.41 & 11.95 &  8.60 &  2.79 &  2.20 &  2.11 &  2.08 &  2.08 & 12.17 &  1.65 \\
 33.5 & 21.17 & 16.19 & 13.43 & 11.82 &  3.59 &  2.53 &  2.22 &  2.15 &  2.13 &  2.12 & 11.96 &  1.66 \\
 34.0 & 23.15 & 17.25 & 14.18 & 12.46 & 11.09 &  3.74 &  2.33 &  2.18 &  2.13 &  2.12 & 12.66 &  1.66 \\
 34.5 & 23.41 & 17.73 & 14.65 & 12.88 & 11.29 &  3.28 &  2.38 &  2.25 &  2.21 &  2.20 & 13.06 &  1.67 \\
 35.0 & 23.51 & 18.05 & 14.84 & 13.08 & 11.12 &  3.17 &  2.43 &  2.29 &  2.25 &  2.24 & 13.37 &  1.68 \\
 36.0 & 26.27 & 19.60 & 16.27 & 14.12 & 12.95 & 10.04 &  2.86 &  2.47 &  2.35 &  2.33 & 14.43 &  1.82 \\
 37.0 & 27.91 & 20.61 & 17.19 & 14.85 & 13.59 & 11.58 &  3.02 &  2.58 &  2.31 &  2.26 & 15.09 &  1.85 \\
 38.0 & 20.96 & 16.24 & 14.08 & 11.43 &  3.99 &  3.16 &  2.54 &  2.32 &  2.24 &  2.23 & 11.95 &  1.68 \\
 39.0 & 14.27 & 11.54 &  8.06 &  6.24 &  4.61 &  3.52 &  2.71 &  2.40 &  2.25 &  2.24 &  6.75 &  1.79 \\
 40.0 & 32.12 & 24.05 & 20.64 & 18.16 & 16.48 & 15.29 &  6.15 &  3.38 &  2.40 &  2.19 & 18.32 &  1.92 \\
 41.0 & 25.86 & 20.31 & 17.51 & 15.90 & 13.73 &  5.41 &  3.73 &  3.09 &  2.44 &  2.29 & 16.02 &  1.88 \\
 42.0 & 31.18 & 23.52 & 20.18 & 18.14 & 16.53 & 15.01 &  4.04 &  3.35 &  2.48 &  2.27 & 18.15 &  1.96 \\
 43.0 & 34.27 & 26.63 & 23.09 & 20.69 & 18.85 & 17.56 & 12.69 &  4.25 &  2.45 &  2.00 & 20.85 &  1.79 \\
 44.0 & 35.94 & 28.69 & 24.78 & 22.23 & 20.40 & 18.98 & 16.07 &  5.23 &  2.60 &  1.66 & 22.22 &  1.66 \\
 45.0 & 16.94 & 12.90 & 11.14 &  9.08 &  7.47 &  6.31 &  4.46 &  3.57 &  2.52 &  2.23 &  9.41 &  1.95 \\
 50.0 & 15.46 & 13.94 & 12.69 & 11.67 & 10.66 &  9.20 &  5.85 &  3.64 &  2.47 &  2.36 & 11.89 &  1.86 \\
 55.0 & 18.14 & 16.46 & 14.50 & 12.60 & 10.95 &  9.50 &  7.21 &  5.73 &  3.39 &  1.96 & 12.59 &  1.96 \\
 60.0 & 43.32 & 34.58 & 29.61 & 27.46 & 25.88 & 24.62 & 11.75 &  9.35 &  4.56 &  1.93 & 27.45 &  1.93 \\
 65.0 & 24.00 & 22.80 & 21.57 & 19.95 & 18.22 & 16.44 & 13.20 & 10.50 &  4.89 &  1.99 & 20.01 &  1.99 \\
 70.0 & 52.95 & 45.53 & 38.23 & 35.35 & 33.35 & 31.71 & 29.03 & 14.53 &  6.57 &  2.21 & 35.21 &  2.08 \\
 75.0 & 28.07 & 26.96 & 25.96 & 24.63 & 23.25 & 21.55 & 18.42 & 15.60 &  8.88 &  2.25 & 24.62 &  2.20 \\
 80.0 & 29.61 & 29.41 & 27.99 & 27.09 & 25.99 & 24.49 & 21.39 & 18.52 & 11.14 &  2.40 & 27.06 &  2.32 \\
 85.0 & 27.68 & 27.69 & 27.62 & 27.23 & 26.17 & 24.93 & 22.41 & 19.93 & 13.28 &  5.15 & 27.32 &  4.44 \\
 90.0 & 26.83 & 26.86 & 26.77 & 26.83 & 26.77 & 26.25 & 24.41 & 21.92 & 15.26 &  7.02 & 27.41 &  6.90 \\
 95.0 & 29.06 & 29.04 & 28.30 & 26.87 & 25.48 & 24.10 & 21.23 & 18.49 & 12.10 &  3.24 & 26.32 &  3.30 \\
100.0 & 40.01 & 38.34 & 36.95 & 35.70 & 34.12 & 31.79 & 25.08 & 13.35 &  2.12 &  2.03 & 35.78 &  1.53 \\
\noalign{\smallskip}
\hline
\end{tabular}
}
\end{table}

% ============
%   Z=solar Remnat Data
% ============
\begin{table}
\caption{$Z=\mathrm{solar}$ Baryonic Remnant Masses \lTab{remnant_soll03}}
\begin{tabular}{rrrrr}
\hline
\hline
\noalign{\smallskip}
Run & SA & SB & SC & SD \\
Energy (\B) & 1.2 & 2.4 & 1.2 & 2.4 \\
Piston & $S=4$ & $S=4$ & Fe core & Fe core \\
\noalign{\smallskip}
\hline
\noalign{\smallskip}
Initial Mass & \multicolumn{4}{c}{\ \hrulefill\ Remnant Mass\ \hrulefill\ \ }\\
(\Msun) & (\Msun) & (\Msun) & (\Msun) & (\Msun) \\
\noalign{\smallskip}
\hline
\noalign{\smallskip}
 12.0 &  1.53 &  1.52 &  1.37 &  1.37 \\
 13.0 &  1.56 &  1.55 &  1.48 &  1.41 \\
 14.0 &  1.71 &  1.70 &  1.57 &  1.52 \\
 15.0 &  1.84 &  1.83 &  1.58 &  1.49 \\
 16.0 &  2.09 &  1.50 &  1.46 &  1.39 \\
 17.0 &  1.54 &  1.54 &  1.52 &  1.42 \\
 18.0 &  1.90 &  1.89 &  1.89 &  1.54 \\
 19.0 &  1.66 &  1.64 &  1.71 &  1.49 \\
 20.0 &  1.86 &  1.82 &  1.96 &  1.62 \\
 21.0 &  1.48 &  1.46 &  1.48 &  1.46 \\
 22.0 &  1.93 &  1.84 &  2.13 &  1.67 \\
 23.0 &  2.36 &  2.14 &  2.75 &  1.95 \\
 24.0 &  2.29 &  2.06 &  2.64 &  1.89 \\
 25.0 &  2.09 &  1.91 &  2.43 &  1.81 \\
 26.0 &  1.75 &  1.74 &  1.82 &  1.61 \\
 27.0 &  1.82 &  1.75 &  1.96 &  1.62 \\
 28.0 &  2.39 &  1.59 &  2.49 &  1.59 \\
 29.0 &  1.76 &  1.64 &  2.02 &  1.57 \\
 30.0 &  1.95 &  1.74 &  2.23 &  1.68 \\
 31.0 &  1.96 &  1.71 &  2.33 &  1.67 \\
 32.0 &  2.27 &  1.79 &  2.62 &  1.79 \\
 33.0 &  2.52 &  1.85 &  2.89 &  1.87 \\
 35.0 &  3.21 &  2.02 &  3.85 &  2.13 \\
 40.0 &  5.60 &  2.73 &  6.72 &  3.15 \\
 45.0 &  3.93 &  2.45 &  5.03 &  2.70 \\
 50.0 &  1.88 &  1.71 &  2.22 &  1.64 \\
 55.0 &  1.76 &  1.66 &  2.05 &  1.57 \\
 60.0 &  1.64 &  1.60 &  1.71 &  1.51 \\
 70.0 &  2.06 &  1.74 &  2.18 &  1.72 \\
 80.0 &  2.03 &  1.67 &  2.05 &  1.65 \\
100.0 &  2.08 &  1.85 &  2.16 &  1.75 \\
\noalign{\smallskip}
\hline
\end{tabular}
\end{table}

\begin{table}
\caption{Remnant Mass Averages and Distributions \lTab{rem}}
\scalebox{0.8}{
\begin{tabular}{rrr|rrrr}
\hline
\hline
\noalign{\smallskip}
\noalign{\smallskip}
Z & piston & $E_\mathrm{exp}$ & BH & $\log(M_\mathrm{BH})$ & BH mass & NS mass \\
  &        & (B)              &(\%)& $(\Msun)$             & (\Msun) & (\Msun) \\
\noalign{\smallskip}
\hline
\noalign{\smallskip}
\multicolumn{7}{c}{assume maximum neutron star gravitational mass of $1.7\,\Msun$} \\
\noalign{\smallskip}
\hline
\noalign{\smallskip}
   solar & S=4 &  1.2 & 23.96 & $ 0.41\pm 0.14$ & $ 2.71\pm 1.02$ & $ 1.41\pm 0.15$ \\
   solar & S=4 &  2.4 & 10.63 & $ 0.35\pm 0.05$ & $ 2.25\pm 0.25$ & $ 1.40\pm 0.13$ \\
   solar &  Fe &  1.2 & 25.48 & $ 0.45\pm 0.16$ & $ 3.06\pm 1.33$ & $ 1.34\pm 0.14$ \\
   solar &  Fe &  2.4 &  7.15 & $ 0.41\pm 0.07$ & $ 2.57\pm 0.38$ & $ 1.33\pm 0.12$ \\
       0 & S=4 &  0.3 & 75.09 & $ 0.86\pm 0.38$ & $10.66\pm 9.64$ & $ 1.39\pm 0.15$ \\
       0 & S=4 &  0.6 & 70.39 & $ 0.79\pm 0.36$ & $ 8.86\pm 8.26$ & $ 1.32\pm 0.14$ \\
       0 & S=4 &  0.9 & 60.25 & $ 0.80\pm 0.33$ & $ 8.66\pm 7.54$ & $ 1.33\pm 0.14$ \\
       0 & S=4 &  1.2 & 52.63 & $ 0.75\pm 0.35$ & $ 7.93\pm 7.47$ & $ 1.33\pm 0.14$ \\
       0 & S=4 &  1.5 & 34.85 & $ 0.76\pm 0.38$ & $ 8.60\pm 8.04$ & $ 1.36\pm 0.15$ \\
       0 & S=4 &  1.8 & 26.31 & $ 0.76\pm 0.40$ & $ 8.80\pm 8.24$ & $ 1.35\pm 0.13$ \\
       0 & S=4 &  2.4 & 19.59 & $ 0.72\pm 0.38$ & $ 7.88\pm 7.39$ & $ 1.35\pm 0.13$ \\
       0 & S=4 &  3.0 & 19.36 & $ 0.63\pm 0.33$ & $ 5.91\pm 5.51$ & $ 1.35\pm 0.12$ \\
       0 & S=4 &  5.0 & 18.89 & $ 0.50\pm 0.24$ & $ 3.85\pm 3.10$ & $ 1.35\pm 0.12$ \\
       0 & S=4 & 10.0 & 17.55 & $ 0.36\pm 0.10$ & $ 2.37\pm 0.79$ & $ 1.36\pm 0.13$ \\
       0 &  Fe &  1.2 & 59.00 & $ 0.74\pm 0.34$ & $ 7.63\pm 7.18$ & $ 1.28\pm 0.19$ \\
       0 &  Fe & 10.0 &  5.23 & $ 0.39\pm 0.16$ & $ 2.67\pm 1.26$ & $ 1.27\pm 0.15$ \\
\noalign{\smallskip}
\hline
\noalign{\smallskip}
\multicolumn{7}{c}{assume maximum neutron star gravitational mass of $2.0\,\Msun$} \\
\noalign{\smallskip}
\hline
\noalign{\smallskip}
   solar & S=4 &  1.2 &  8.59 & $ 0.56\pm 0.12$ & $ 3.80\pm 1.02$ & $ 1.47\pm 0.21$ \\
   solar & S=4 &  2.4 &  3.72 & $ 0.41\pm 0.02$ & $ 2.56\pm 0.11$ & $ 1.43\pm 0.17$ \\
   solar &  Fe &  1.2 & 14.53 & $ 0.55\pm 0.15$ & $ 3.76\pm 1.40$ & $ 1.40\pm 0.22$ \\
   solar &  Fe &  2.4 &  4.77 & $ 0.45\pm 0.04$ & $ 2.80\pm 0.23$ & $ 1.34\pm 0.15$ \\
       0 & S=4 &  0.3 & 70.44 & $ 0.90\pm 0.36$ & $11.22\pm 9.68$ & $ 1.45\pm 0.20$ \\
       0 & S=4 &  0.6 & 60.26 & $ 0.87\pm 0.33$ & $10.00\pm 8.41$ & $ 1.45\pm 0.25$ \\
       0 & S=4 &  0.9 & 56.26 & $ 0.83\pm 0.32$ & $ 9.11\pm 7.60$ & $ 1.38\pm 0.22$ \\
       0 & S=4 &  1.2 & 47.74 & $ 0.79\pm 0.33$ & $ 8.52\pm 7.59$ & $ 1.37\pm 0.20$ \\
       0 & S=4 &  1.5 & 30.18 & $ 0.83\pm 0.36$ & $ 9.61\pm 8.19$ & $ 1.39\pm 0.18$ \\
       0 & S=4 &  1.8 & 21.93 & $ 0.84\pm 0.38$ & $10.14\pm 8.41$ & $ 1.38\pm 0.17$ \\
       0 & S=4 &  2.4 & 14.16 & $ 0.88\pm 0.34$ & $10.08\pm 7.62$ & $ 1.39\pm 0.17$ \\
       0 & S=4 &  3.0 & 13.15 & $ 0.77\pm 0.31$ & $ 7.70\pm 5.89$ & $ 1.39\pm 0.17$ \\
       0 & S=4 &  5.0 & 10.85 & $ 0.62\pm 0.26$ & $ 5.11\pm 3.62$ & $ 1.40\pm 0.19$ \\
       0 & S=4 & 10.0 &  1.79 & $ 0.59\pm 0.16$ & $ 4.16\pm 1.55$ & $ 1.44\pm 0.22$ \\
       0 &  Fe &  1.2 & 51.83 & $ 0.79\pm 0.32$ & $ 8.39\pm 7.35$ & $ 1.37\pm 0.27$ \\
       0 &  Fe & 10.0 &  1.38 & $ 0.62\pm 0.14$ & $ 4.41\pm 1.34$ & $ 1.29\pm 0.18$ \\
\noalign{\smallskip}
\hline
\end{tabular}
}
\end{table}

\begin{table}
\caption{Remnant Mass Averages and Distributions by Origin \lTab{rem_org}}
\scalebox{0.8}{
\begin{tabular}{rrrrrr}
\hline
\hline
\noalign{\smallskip}
\noalign{\smallskip}
range    &  BH &  NS  & $\log(M_\mathrm{BH})$ & BH mass & NS mass \\
$(\Msun)$& (\%)& (\%) & $(\Msun)$             & (\Msun) & (\Msun) \\
\noalign{\smallskip}
\hline
\noalign{\smallskip}
\multicolumn{6}{c}{Case:\ \ $Z=\mathrm{solar}$,\ \ $E= 1.2\,$B,\ \ piston at $S=4$,\ \ $M_{\mathrm{NS}}^{\mathrm{max}}=1.7\,\Msun$} \\
\noalign{\smallskip}
\hline
\noalign{\smallskip}
$ < 10$ &    -- & 12.44 &              -- &              -- & $ 1.24\pm 0.00$ \\
$10-12$ &    -- & 20.00 &              -- &              -- & $ 1.27\pm 0.04$ \\
$12-15$ &    -- & 18.64 &              -- &              -- & $ 1.46\pm 0.08$ \\
$15-20$ &  3.52 & 13.54 & $ 0.30\pm 0.01$ & $ 2.02\pm 0.04$ & $ 1.56\pm 0.08$ \\
$20-25$ &  4.93 &  4.42 & $ 0.35\pm 0.02$ & $ 2.22\pm 0.11$ & $ 1.50\pm 0.10$ \\
$25-40$ &  8.98 &  3.52 & $ 0.44\pm 0.14$ & $ 2.93\pm 1.06$ & $ 1.61\pm 0.04$ \\
$ > 40$ &  6.53 &  3.48 & $ 0.46\pm 0.17$ & $ 3.14\pm 1.25$ & $ 1.57\pm 0.06$ \\
\hline
  total & 23.96 & 76.04 & $ 0.41\pm 0.14$ & $ 2.71\pm 1.02$ & $ 1.41\pm 0.15$ \\
\noalign{\smallskip}
\hline
\noalign{\smallskip}
\multicolumn{6}{c}{Case:\ \ $Z=\mathrm{solar}$,\ \ $E= 1.2\,$B,\ \ piston at $S=4$,\ \ $M_{\mathrm{NS}}^{\mathrm{max}}=2.0\,\Msun$} \\
\noalign{\smallskip}
\hline
\noalign{\smallskip}
$ < 10$ &    -- & 12.44 &              -- &              -- & $ 1.24\pm 0.00$ \\
$10-12$ &    -- & 20.00 &              -- &              -- & $ 1.27\pm 0.04$ \\
$12-15$ &    -- & 18.64 &              -- &              -- & $ 1.46\pm 0.08$ \\
$15-20$ &    -- & 17.06 &              -- &              -- & $ 1.60\pm 0.11$ \\
$20-25$ &  0.39 &  8.96 & $ 0.37\pm 0.00$ & $ 2.36\pm 0.01$ & $ 1.70\pm 0.22$ \\
$25-40$ &  4.86 &  7.63 & $ 0.55\pm 0.12$ & $ 3.64\pm 0.98$ & $ 1.72\pm 0.12$ \\
$ > 40$ &  3.34 &  6.67 & $ 0.61\pm 0.10$ & $ 4.18\pm 0.92$ & $ 1.67\pm 0.12$ \\
\hline
  total &  8.59 & 91.41 & $ 0.56\pm 0.12$ & $ 3.80\pm 1.02$ & $ 1.47\pm 0.21$ \\
\noalign{\smallskip}
\hline
\noalign{\smallskip}
\multicolumn{6}{c}{Case:\ \ $Z=\mathrm{0}$,\ \ $E= 1.2\,$B,\ \ piston at $S=4$,\ \ $M_{\mathrm{NS}}^{\mathrm{max}}=1.7\,\Msun$} \\
\noalign{\smallskip}
\hline
\noalign{\smallskip}
$ < 10$ &    -- &  6.98 &              -- &              -- & $ 1.16\pm 0.00$ \\
$10-12$ &    -- & 21.24 &              -- &              -- & $ 1.26\pm 0.05$ \\
$12-15$ &  3.20 & 16.60 & $ 0.34\pm 0.03$ & $ 2.22\pm 0.16$ & $ 1.44\pm 0.13$ \\
$15-20$ & 15.62 &  2.50 & $ 0.50\pm 0.07$ & $ 3.23\pm 0.50$ & $ 1.56\pm 0.05$ \\
$20-25$ &  9.94 &    -- & $ 0.62\pm 0.12$ & $ 4.28\pm 1.21$ &              -- \\ 
$25-40$ & 13.24 &  0.04 & $ 0.81\pm 0.30$ & $ 7.95\pm 4.42$ & $ 1.69\pm 0.00$ \\
$ > 40$ & 10.63 &    -- & $ 1.27\pm 0.17$ & $19.91\pm 7.18$ &              -- \\ 
\hline
  total & 52.63 & 47.37 & $ 0.75\pm 0.35$ & $ 7.93\pm 7.47$ & $ 1.33\pm 0.14$ \\
\noalign{\smallskip}
\hline
\noalign{\smallskip}
\multicolumn{6}{c}{Case:\ \ $Z=\mathrm{0}$,\ \ $E= 1.2\,$B,\ \ piston at $S=4$,\ \ $M_{\mathrm{NS}}^{\mathrm{max}}=2.0\,\Msun$} \\
\noalign{\smallskip}
\hline
\noalign{\smallskip}
$ < 10$ &    -- &  6.98 &              -- &              -- & $ 1.16\pm 0.00$ \\
$10-12$ &    -- & 21.24 &              -- &              -- & $ 1.26\pm 0.05$ \\
$12-15$ &  1.19 & 18.61 & $ 0.38\pm 0.00$ & $ 2.38\pm 0.02$ & $ 1.48\pm 0.17$ \\
$15-20$ & 14.43 &  3.70 & $ 0.52\pm 0.05$ & $ 3.32\pm 0.40$ & $ 1.65\pm 0.15$ \\
$20-25$ &  9.94 &    -- & $ 0.62\pm 0.12$ & $ 4.28\pm 1.21$ &              -- \\ 
$25-40$ & 11.55 &  1.73 & $ 0.88\pm 0.25$ & $ 8.81\pm 4.08$ & $ 1.82\pm 0.08$ \\
$ > 40$ & 10.63 &    -- & $ 1.27\pm 0.17$ & $19.91\pm 7.18$ &              -- \\ 
\hline
  total & 47.74 & 52.26 & $ 0.79\pm 0.33$ & $ 8.52\pm 7.59$ & $ 1.37\pm 0.20$ \\
\noalign{\smallskip}
\hline
\end{tabular}
}
\end{table}

\clearpage

%%%% figures

% comparison of kepler and pangu hydro at different times in model Z25D 
\begin{figure}
\epsscale{0.75}
%\plotone{comp.eps}
\plotone{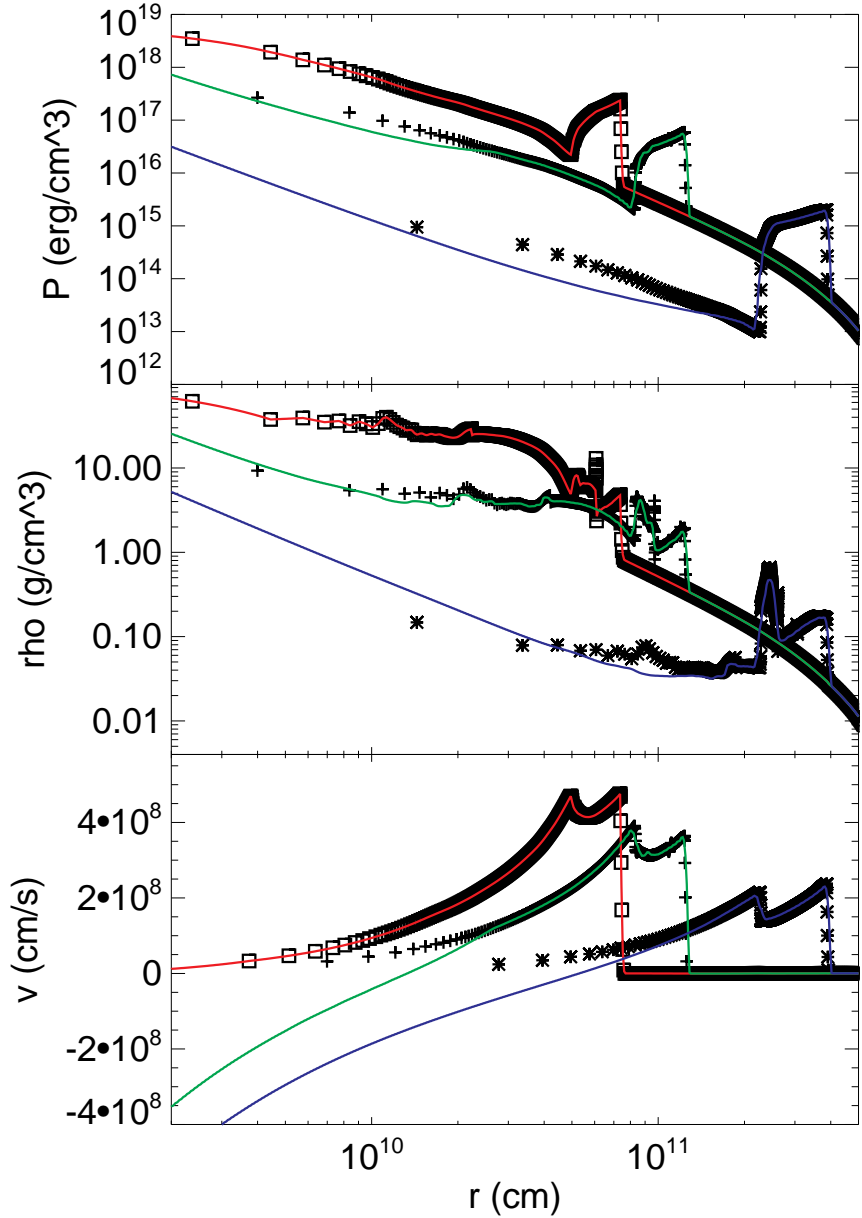}
\caption{Pressure, density and velocity profiles at 100 (red lines and
  square symbols), 200 (green lines and plus symbols), and 1000\,s
  (blue lines and star symbols) in Model Z25D calculated using Kepler
  (symbols) and Pangu (solid lines). The agreement is excellent except
  near the origin. Since Pangu uses a more realistic representation of
  the fallback at small radii, its results are preferred.  The inner
  boundary in Pangu is inside the sonic radius at all times
  \lFig{comphydro}}
\end{figure}

% mdot Z25D, S25A
\begin{figure}
\epsscale{0.8}
%\plotone{mmdot.eps}
\plotone{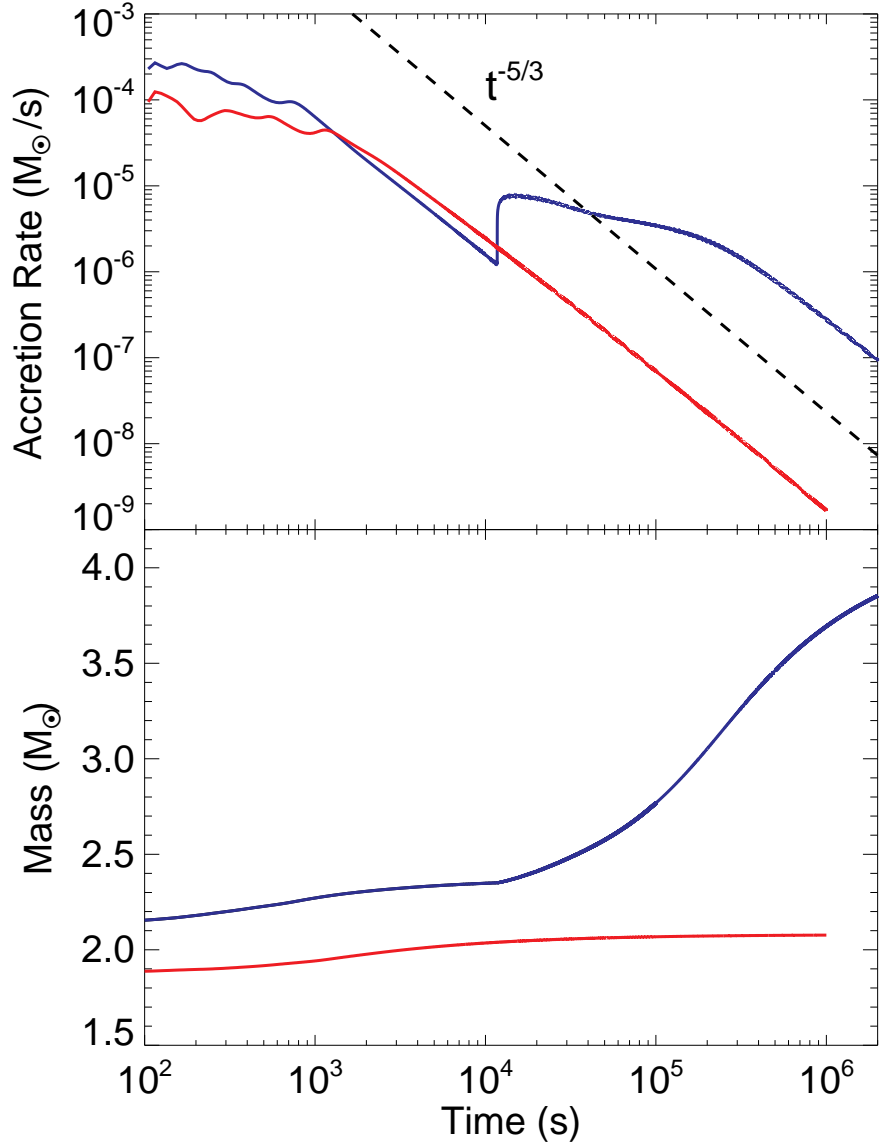}
\caption{Accretion rates and central point mass for models Z25D (blue
  lines) and S25A (red lines). The dotted line shows the asymptotic
  accretion rate, $\sim t^{-5/3}$. Note the prominent appearance of
  the reverse shock at the core at about $10^4$\,s in Z25D.  For model
  S25A the reverse shock has not arrived back at the origin at
  10$^6$\,s and, in fact, is still moving outwards in space. Its
  eventual arrival will have little consequence for the mass of the
  remnant. Note a period of about 1000\,s during which the initial
  accretion rate is nearly constant.  \lFig{mdot25}}
\end{figure}

% Z25B,D,G all mass
\begin{figure}
\epsscale{0.8}
%\plotone{mm_BDG_pangu.eps}
\plotone{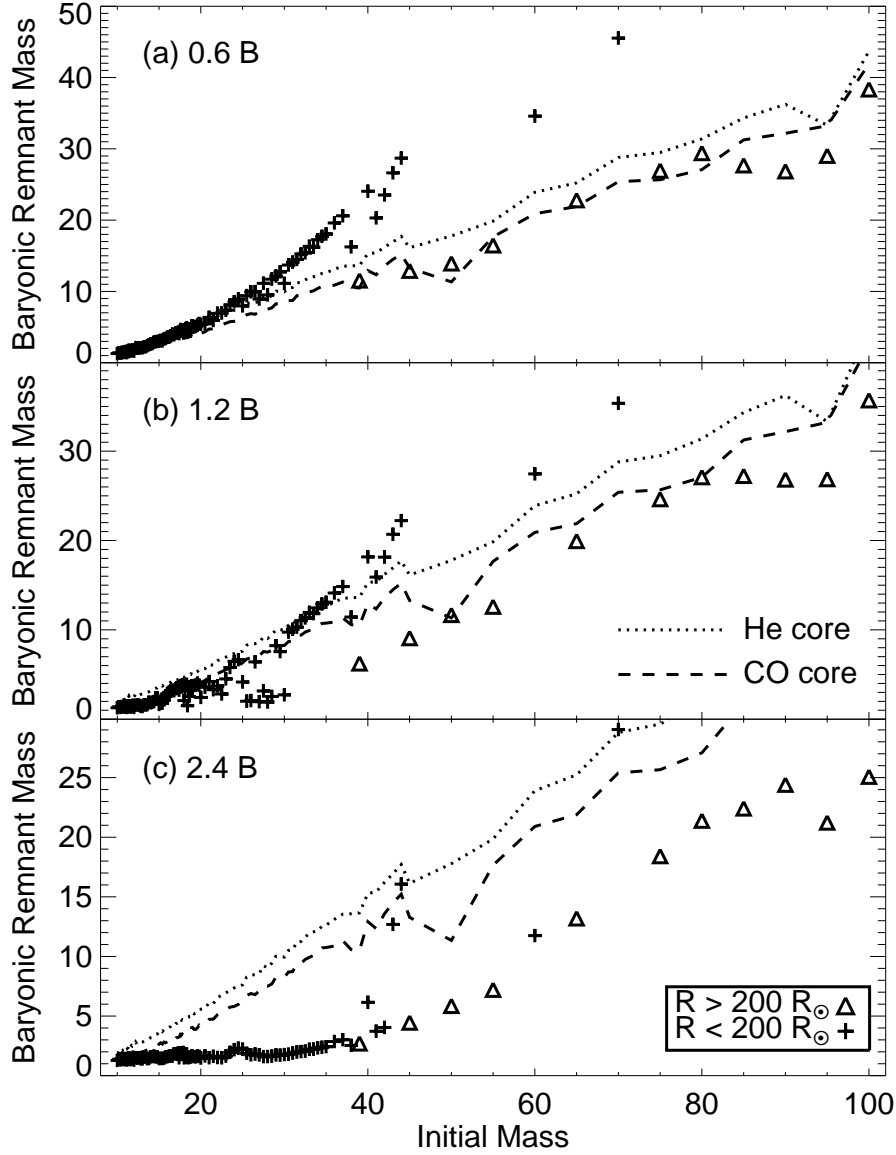}
\caption{Comparison of baryonic remnant masses for (a) ZB, (b) ZD and (c) ZG
  models.  The explosion energies are 0.6, 1.2 and 2.4\,B, for models
  ZB, ZD and ZG, respectively. It is clear that there are two branches
  of remnant masses.  The higher mass branch consists of compact stars
  with a radius less than $200\,R_\odot$, whereas the lower mass
  branch consists of red supergiants with a radius greater than
  $200\,R_\odot$.  The positions of the He core (dotted lines) and CO
  core (dashed lines) in the initial models are also shown.  Note that
  for the lower mass branch of ZB models the remnant mass is very
  close to the CO core mass.  \lFig{remnantz}}
\end{figure}

% Z25B,D,G low mass
\begin{figure}
\epsscale{0.8}
%\plotone{mm_BDG_pangu_sml.eps}
\plotone{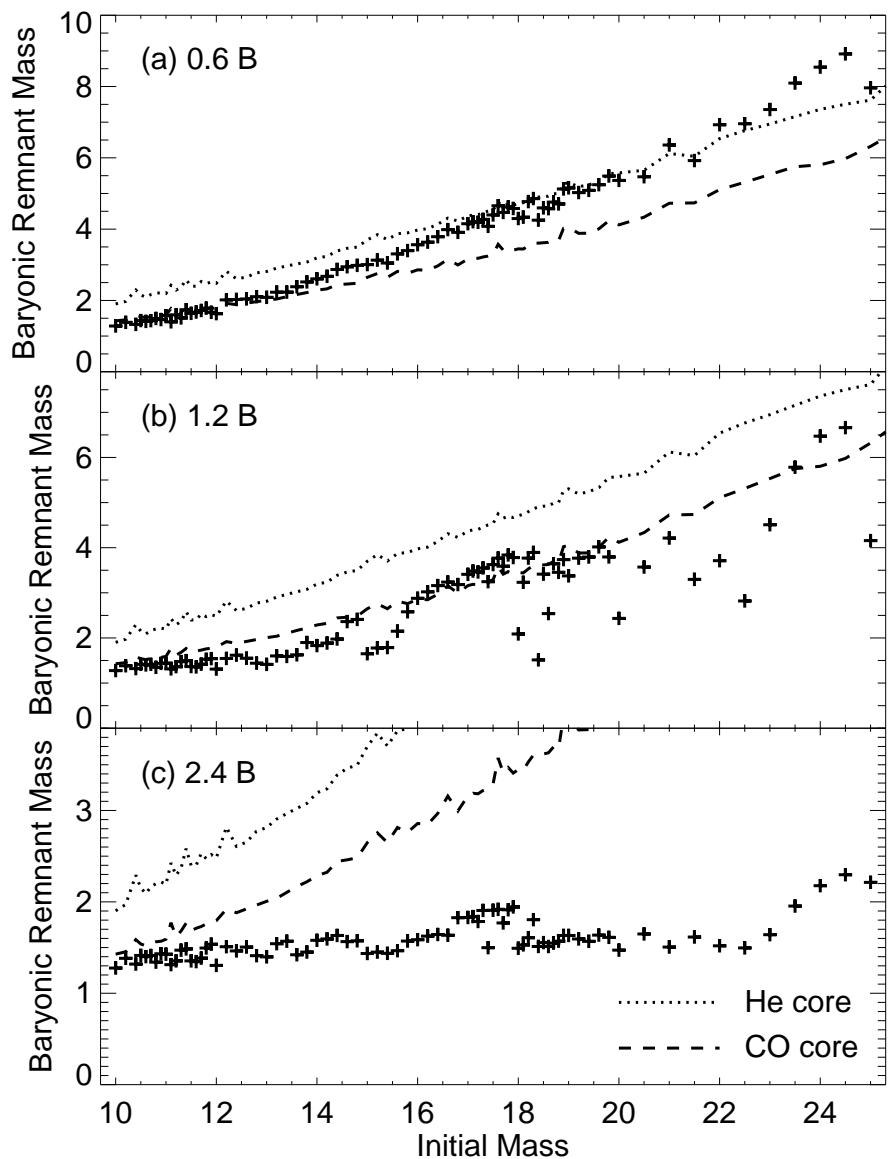}
\caption{Baryonic remnant masses for (a) ZB, (b) ZD and (c) ZG models
  plotted on a finer scale for lower mass stars.  The explosion
  energies are 0.6, 1.2 and 2.4\,B, for models ZB, ZD and ZG,
  respectively.  As we expected, ZG models make many neutron stars,
  whereas the lower energy ZB models make many black holes.
  \lFig{remlow}}
\end{figure}

% neutron star imf - S=4 piston mass, Z=0, 0.6 B
\begin{figure}
\includegraphics[angle=90,width=\columnwidth]{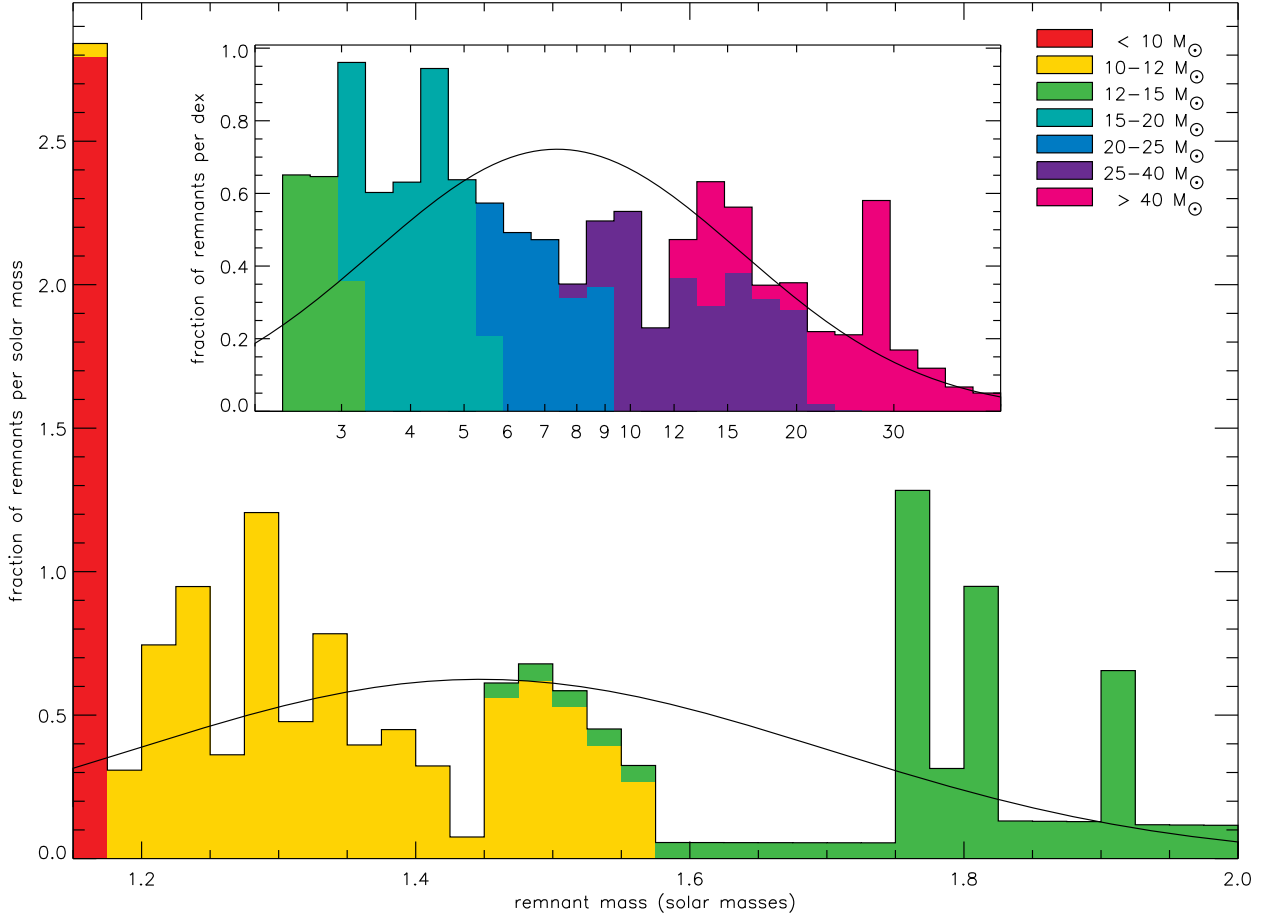}
\caption{Distribution of remnant masses for 0.6\,B explosions of
  metal-free stars with pistons located at the $S/N_Ak = 4.0$ point,
  for an initial mass range of $9.5\,\Msun$ to $100\,\Msun$, and an
  assumed a maximum neutron stars mass of $2\,\Msun$.  The main figure
  gives \emph{gravitational} masses of neutron stars, the insert shows
  the \emph{baryonic} masses of black holes.  The color coding
  (cumulative) indicates the initial mass range of the progenitor
  stars.  The curve is a Gaussian fit with the same average and
  variance as distribution for the neutron stars (main figure).  For
  the insert the curve is a Gaussian fit to the logarithm of black
  hole masses (geometric fit).  The normalization of the bins in the
  big plot is such that the sum over the bins times the bin width
  equals total fraction of neutron stars.  For the inserts the
  normalization is not ``per solar mass'' but ``per dex'', i.e., the
  sum of bin height times bin width in dex equals the total fraction
  of black holes.  The bin sizes are $0.025\,\Msun$ for the main
  figure and $0.05\,$dex for the insert. The spike at 1.18 \Msun \ is
  for lower mass stars which make iron cores near the Chandrasekhar
  mass limit and have final mass cuts near the boundary of that
  core. Since iron cores have an appreciable neutron excess, the
  Chandrasekhar mass is appreciably reduced from the classical 1.39
  \Msun \ of baryons (1.26 \Msun \ gravitational).
 \lFig{nstar04-0.6B}}
\end{figure}

% neutron star imf - S=4 piston mass, Z=0, 1.2 B
\begin{figure}
\includegraphics[angle=90,width=\columnwidth]{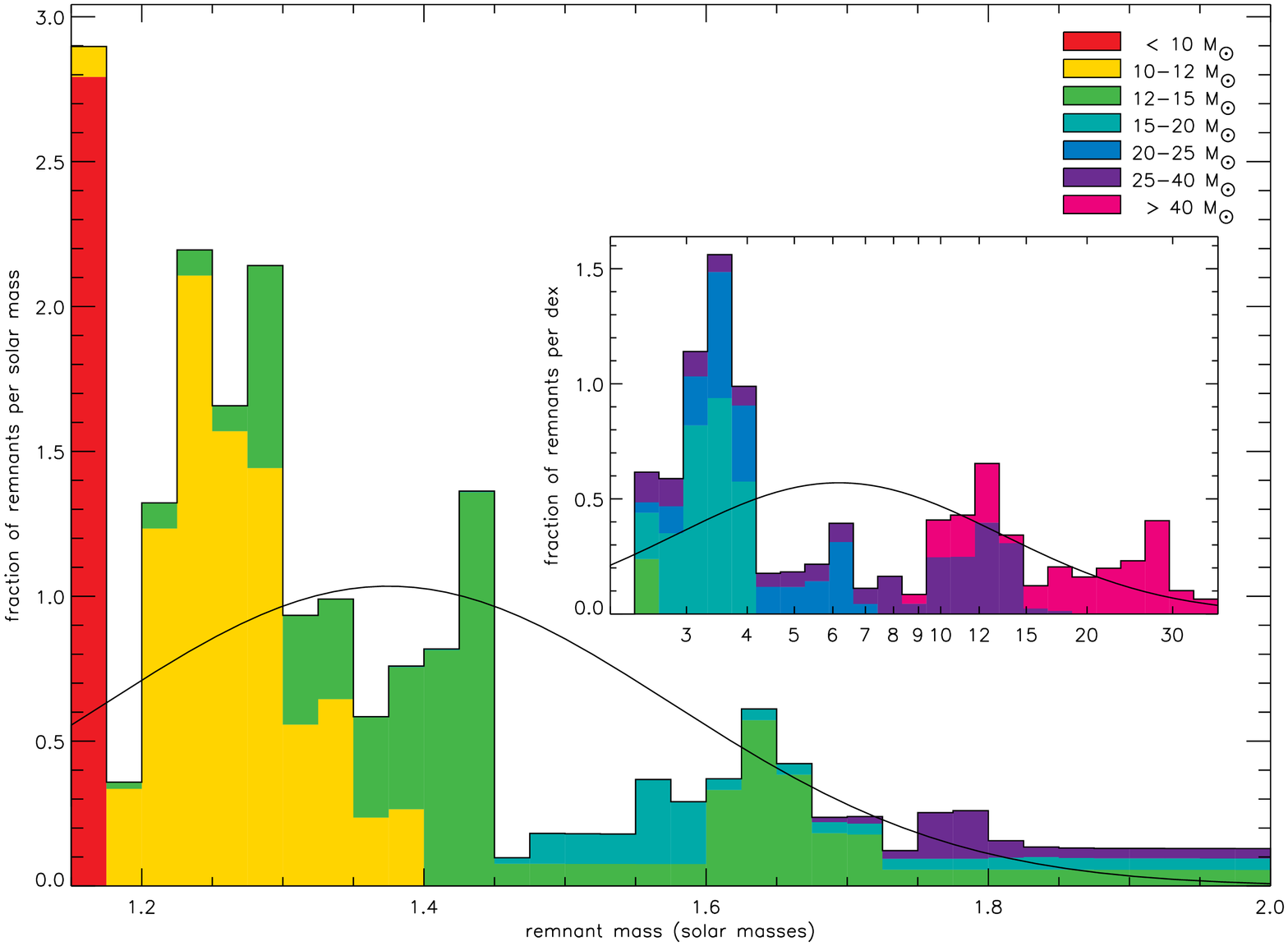}
\caption{ Distribution of remnant masses for 1.2\,B explosions of
  metal-free stars with pistons located at the $S/N_Ak = 4.0$ point,
  for an initial mass range of $9.5\,\Msun$ to $100\,\Msun$, and an
  assumed a maximum neutron stars mass of $2\,\Msun$. See also the
  caption of \Fig{nstar04-0.6B}.  \lFig{nstar04-1.2B}}
\end{figure} 

% neutron star imf - S=4 piston mass, Z=0, 2.4 B
\begin{figure}
\includegraphics[angle=90,width=\columnwidth]{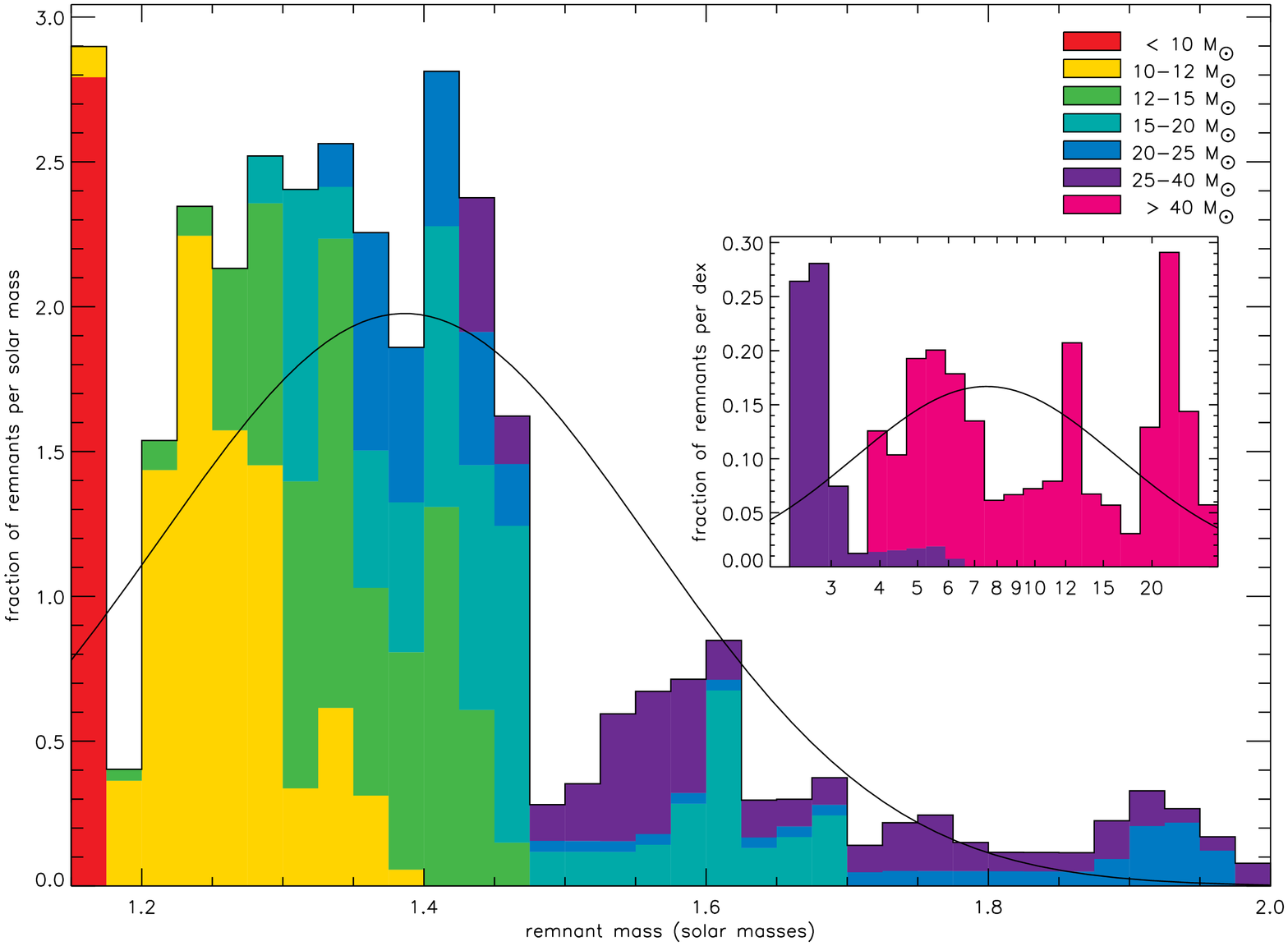}
\caption{ Distribution of remnant masses for 2.4\,B explosions of
  metal-free stars with pistons located at the $S/N_Ak = 4.0$ point,
  for an initial mass range of $9.5\,\Msun$ to $100\,\Msun$, and an
  assumed a maximum neutron stars mass of $2\,\Msun$.  See also the
  caption of \Fig{nstar04-0.6B}.  \lFig{nstar04-2.4B}}
\end{figure} 

% neutron star imf - Fe core piston mass, Z=0, 1.2 B
\begin{figure}
\includegraphics[angle=90,width=\columnwidth]{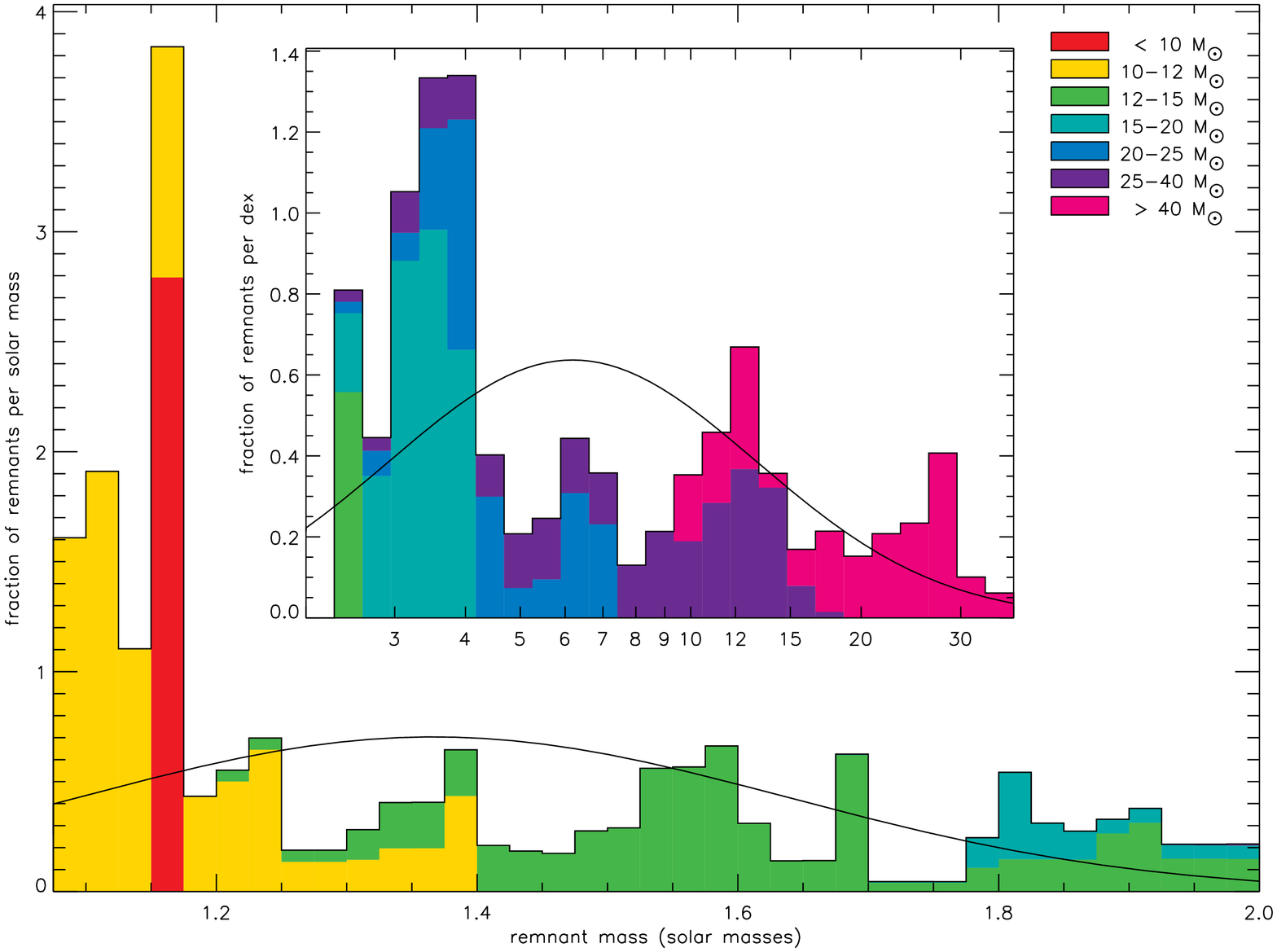}
\caption{Distribution of remnant masses for 1.2\,B explosions of
  metal-free stars with pistons located at the edge of the
  deleptonized core, for an initial mass range of $9.5\,\Msun$ to
  $100\,\Msun$, and an assumed a maximum neutron stars mass of
  $2\,\Msun$.  See also the caption of \Fig{nstar04-0.6B}.
  \lFig{nstar0Fe}}
\end{figure}

% S25A compare with kepler
\begin{figure}
\epsscale{0.9}
%\plotone{mass.sA_pangu.eps}
\plotone{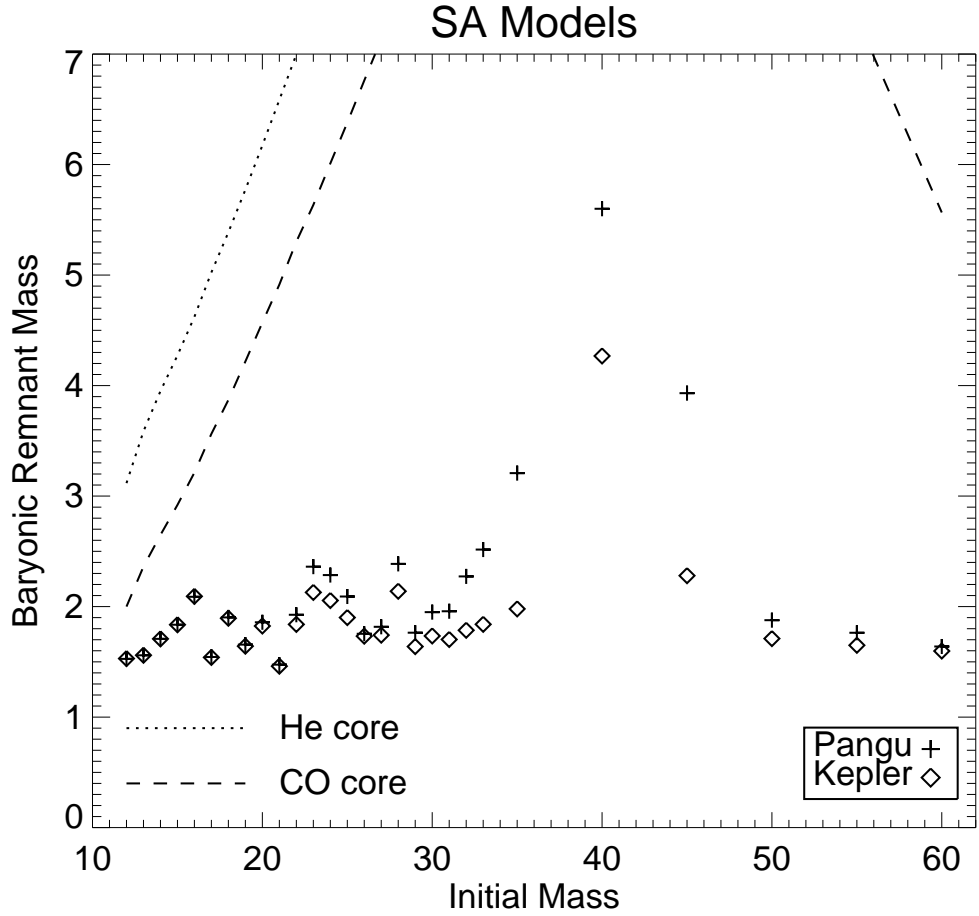}
\caption{Comparison of baryonic remnant masses for SA series with
  Kepler and Pangu.  The results from two different codes are similar.
  However, the final baryonic remnant masses calculated using Pangu
  (plus symbols) are greater than those calculated using Kepler
  (diamond symbols), especially for the initial mass range of
  $30-50\,\Msun$.  \lFig{remnantsol}}
\end{figure}

% neutron star imf - s=4 piston mass, solar, 1.2 B
\begin{figure}
\includegraphics[angle=90,width=\columnwidth]{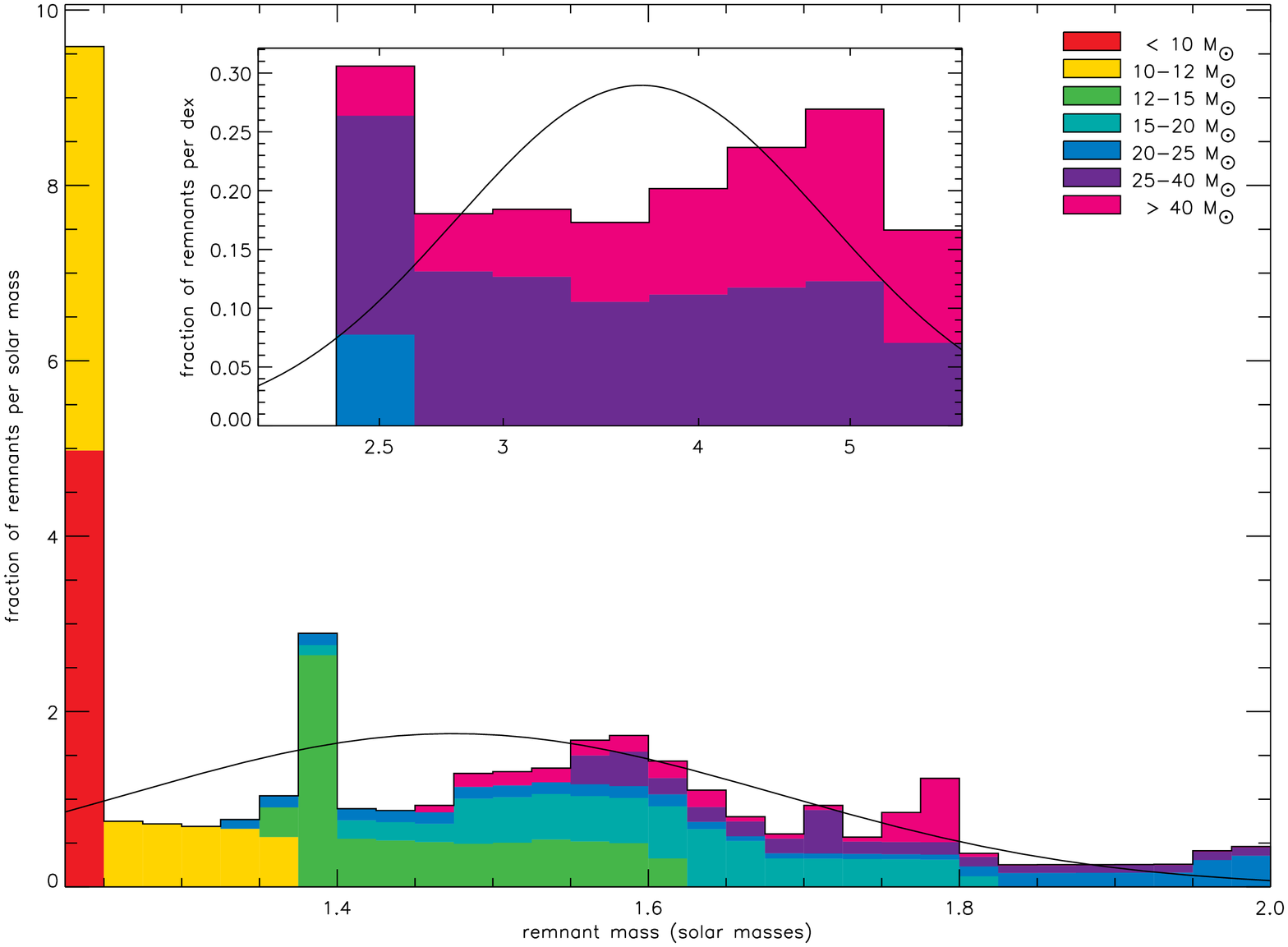}
\caption{Distribution of remnant masses for 1.2\,B explosions of solar
  metallicity stars with pistons located at the entropy $S/N_Ak = 4$
  point for an initial mass range of $9.1\,\Msun$ to $100\,\Msun$.  We
  assumed a maximum neutron stars mass of $2\,\Msun$.  See also the
  caption of \Fig{nstar04-0.6B}.  \lFig{nstarsol4}}
\end{figure} 

% neutron star imf - Fe core piston mass, solar, 1.2 B
\begin{figure}
\includegraphics[angle=90,width=\columnwidth]{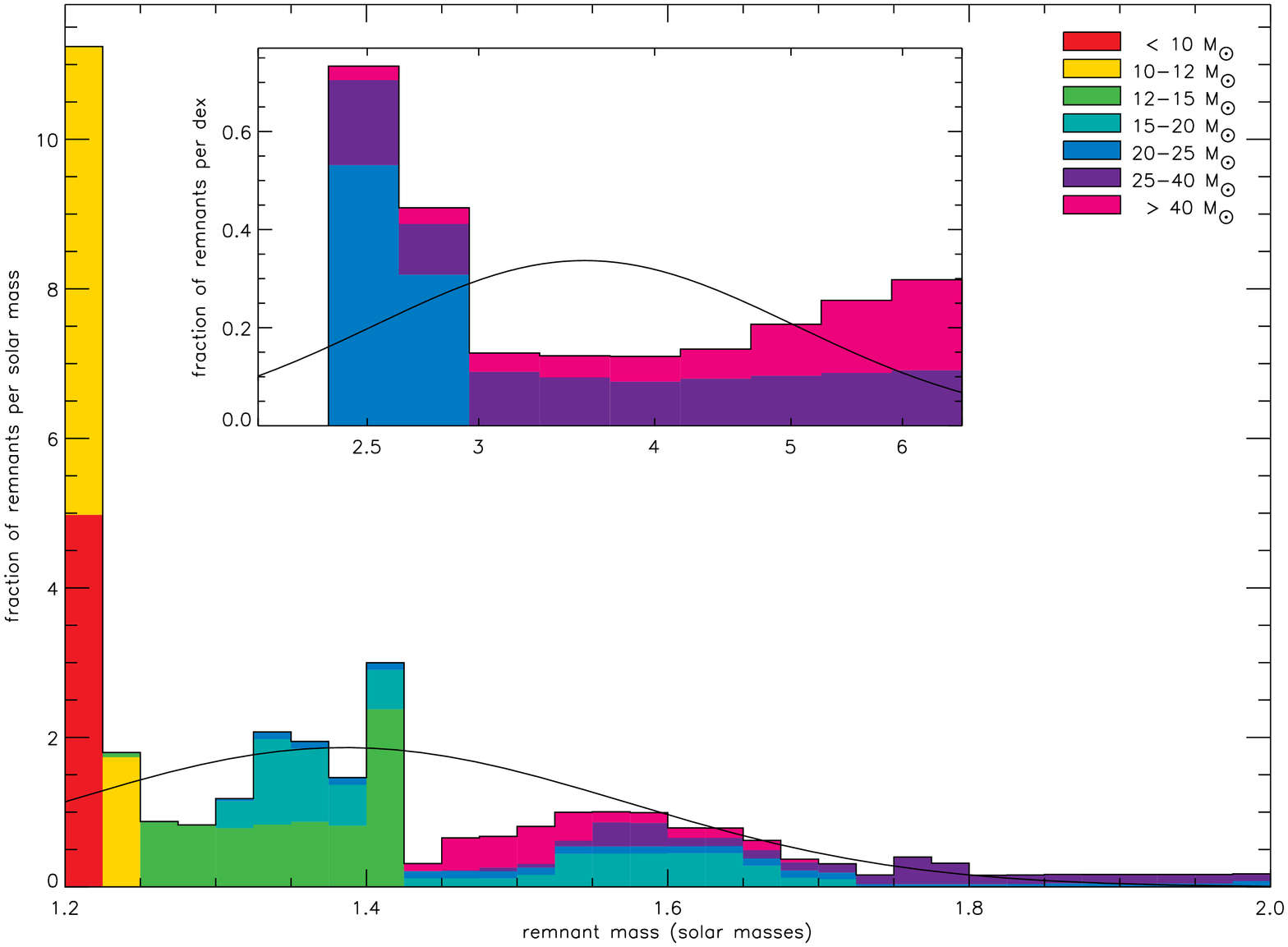}
\caption{Distribution of remnant masses for 1.2\,B explosions of solar
  metallicity stars with pistons located at the edge of the
  deleptonized core, for an initial mass range of $9.1\,\Msun$ to
  $100\,\Msun$, and an assumed maximum neutron star mass of
  $2\,\Msun$.   See also the caption of \Fig{nstar04-0.6B}.
  \lFig{nstarsolFe}}
\end{figure} 

\end{document}